\documentclass[aps,prd,longbibliography,reprint,twocolumn,amsmath,amssymb,amsfonts,showpacs,footnote,superscriptaddress]{revtex4-2}

\usepackage[T1]{fontenc}
\usepackage{lmodern}
\usepackage{amsmath,amssymb,amsfonts}
\usepackage{bm}
\usepackage{mathtools}
\usepackage{physics}
\usepackage{hyperref}
\usepackage{xcolor}
\usepackage[T1]{fontenc}
\usepackage{booktabs}
\usepackage{siunitx}
\usepackage{url}

\usepackage{orcidlink}

\definecolor{navyblue}{rgb}{0.0, 0.0, 0.5}
\definecolor{ferrarired}{rgb}{1.0, 0.11, 0.0}
\definecolor{persianblue}{rgb}{0.11, 0.22, 0.73}
\usepackage{float}

\hypersetup{
	colorlinks=true,
	citecolor=blue,%
	linkcolor=ferrarired,%
	urlcolor=persianblue,%
	filecolor=blue,
	linktoc=page
}

\newcommand{\W}[2]{\mathcal{W}\!\left[#1,#2\right]}

\begin{document}
	
	\title{Scattering from compact objects: Debye series and Regge-Debye poles}
	
	\author{Mohamed \surname{Ould~El~Hadj}\,\orcidlink{0000-0002-8558-7992}}
	\email{med.ouldelhadj@gmail.com}
	
	\date{\today}

\begin{abstract}	
	  	
  	 We investigate elastic scattering by a compact, horizonless body in curved spacetime, considering a massless scalar wave incident on a static, spherically symmetric, uniform-density star of radius $R$ and mass $M$ with a Schwarzschild exterior. We introduce an exact Debye-series decomposition of the scattering matrix, in the spirit of Debye expansions in Mie scattering. This decomposition separates direct surface reflection from contributions involving transmission into the interior and subsequent propagation, and admits a natural trajectory interpretation. We then determine the associated Regge-Debye pole spectrum in the complex angular momentum plane. For neutron star-like tenuities ($R>3M$), the spectrum exhibits two pole families: a surface-wave branch associated with the surface matching condition and a broad-resonance branch associated with the interior regularity condition. For ultracompact objects ($R<3M$), the surface-wave branch persists, while the interior-resonance sector splits into broad- and narrow-resonance branches. 	
  	We next reconstruct the scattering amplitude from the Debye partial-wave contributions and find excellent agreement with direct partial-wave calculations. Finally, we develop complex angular momentum representations order by order in the Debye series, making explicit how the pole families and nonpole sectors contribute to each Debye term. In the neutron star-like regime, we find a genuine competition between Regge-Debye pole sums and branch-cut contributions, and show that, at high frequency, the  rainbow-like enhancement already arises from the first interior-transmission contribution and is dominated by the interior-resonance Regge-Debye poles. By contrast, in the ultracompact regime, the Debye amplitudes are overwhelmingly pole dominated.
\end{abstract}

	\maketitle
	
	\tableofcontents

%=========================================================
\section{Introduction}
\label{sec:intro}

Time-independent wave scattering by compact objects offers a clean arena in which strong-field gravity, wave interference, and, for horizonless bodies, internal structure leave imprints on observable angular spectra. This perspective has acquired renewed relevance in the era of gravitational-wave astronomy \cite{LIGOScientific:2016aoc,LIGOScientific:2017vwq,KAGRA:2021vkt} and horizon-scale imaging \cite{EventHorizonTelescope:2019dse,EventHorizonTelescope:2022xqj}, where distinguishing between black holes and horizonless compact-object scenarios~\cite{Cardoso:2019rvt} motivates a careful characterization of wave propagation in the near-field region. On the theoretical side, most quantitative studies of scattering in general relativity ultimately rely on partial-wave expansions (see, e.g., Ref.~\cite{Futterman:1988ni}), a natural approach on highly symmetric backgrounds. In particular, decades of work on black holes have established a detailed understanding across spins and frequency regimes (see, e.g., Refs.~\cite{Hildreth1964PhDT64,Matzner:1968,Vishveshwara:1970,Futterman:1988ni,Mashhoon:1973zz,Fabbri:1975,Sanchez:1977vz,MatznerRyan1978,Handler:1980un,Matzner:1985rjn,Andersson:1995vi,Glampedakis:2001cx,Dolan:2008kf,Crispino:2009xt} and references therein), including characteristic signatures such as orbiting oscillations and backward glory enhancements (see, e.g., Refs.~\cite{Matzner:1968,Futterman:1988ni,Dolan:2008kf,Crispino:2009xt} and \cite{Stratton:2019deq,Dolan:2017rtj,OuldElHadj:2019kji} for recent developments on rainbow scattering).

Despite their effectiveness, partial-wave expansions have familiar drawbacks. Long-range interactions may produce slow convergence and require explicit acceleration procedures, and the physical content of a multipole sum is often not transparent. Complex angular momentum (CAM) methods offer a natural reorganization of the scattering problem: the partial-wave sum is replaced by a contour integral in the complex angular momentum plane, via the Sommerfeld-Watson transform, whose deformation isolates pole contributions (Regge poles) together with nonpole background terms. Developed originally in diffraction theory~\cite{Watson18,Sommerfeld49}, CAM techniques are now standard in resonant scattering~\cite{deAlfaro:1965zz,Newton:1982qc,Nussenzveig:2006,Grandy2000,Uberall1992,AkiRichards2002,Gribov69,Collins77,BaronePredazzi2002,DonnachieETAL2005}. In black hole physics, they provide a particularly transparent interpretation, with Regge poles encoding surface-wave propagation near the photon sphere and linking naturally to the quasinormal mode spectrum~\cite{Andersson:1994rk,Andersson:1994rm,Decanini:2002ha}. They have also proved useful for accurate reconstructions of scattering observables and for semiclassical descriptions of orbiting, glory, absorption, and radiation phenomena~\cite{Folacci:2019cmc,Folacci:2019vtt,Decanini:2009mu,Decanini:2011xi,Decanini:2011xw,Folacci:2018sef,Torres:2022fyf,OuldElHadj:2025hbl}. By comparison, CAM analyses of horizonless relativistic stars remain relatively limited: an early study by Chandrasekhar and Ferrari related stellar Regge poles to resonant energy transport through the star~\cite{Chandrasekhar449}, while a more recent compact-body scattering analysis was given in Ref.~\cite{OuldElHadj:2019kji}.

By contrast, scattering by a horizonless compact body (with a regular center and radius $R$) is intrinsically richer. This is because waves may penetrate the object and reemerge to interfere with the exterior component. The scattering pattern depends not only on the usual frequency parameter $M\omega$, but also on the compactness (or tenuity) $R/M$ and on the chosen interior model. In particular, compact-body scattering admits regimes in which interior and exterior ray families coexist, producing interference structures without a direct analog in the purely exterior black hole problem~\cite{Dolan:2017rtj,Stratton:2019deq,OuldElHadj:2019kji}. This enhanced sensitivity is directly relevant to current efforts to distinguish black holes from alternative compact-object scenarios, including neutron stars and more speculative ultracompact objects (UCOs)~\cite{Cardoso:2019apo,Maggio:2021ans,DeLaurentis:2025nsy}. See also Refs.~\cite{Nambu:2019sqn,He:2019orl,Alexandre:2018crg,Marchant:2019swq,Seenivasan:2025ysy} for complementary perspectives on wave propagation and related phenomena in compact-body spacetimes.
 
One striking manifestation is \emph{rainbow scattering}. In the short-wavelength regime, compact objects exhibit  rainbow-like enhancements associated with stationary points of the deflection function and the formation of caustics~\cite{Dolan:2017rtj,Stratton:2019deq}. For neutron star-like tenuity ($R/M\sim6$), the corresponding rainbow angles can be large and may depend sensitively on the interior structure, making them, in principle, a diagnostic of the matter distribution.

In this paper, we extend the CAM program by introducing an exact Debye-series expansion of the compact-body $S$-matrix, in the spirit of Debye expansions in Mie scattering~\cite{Nussenzveig:2006}. To our knowledge, this work provides the first application of a Regge-Debye decomposition to wave scattering by compact objects in general relativity. This expansion separates direct surface reflection from the sequence of contributions involving transmission into the interior and subsequent propagation. It admits a trajectory interpretation in which the leading term represents reflection at the surface, while higher-order terms correspond to waves that reemerge after successive surface-center-surface traversals. We then develop CAM representations \emph{order by order} in the Debye series. This yields a refined pole classification in terms of \emph{Regge-Debye} families associated with the surface matching condition and with the interior regularity condition, and it clarifies how each family contributes to specific Debye orders and observable scattering features.

Implementing the formalism for representative neutron star-like and ultracompact configurations, we find that the neutron star-like regime ($R\gtrsim 3M$) displays a genuine interplay between Regge-Debye pole sums and nonpole contributions, whereas in the ultracompact regime ($R<3M$) the Debye amplitudes are overwhelmingly dominated by pole contributions over the angular ranges considered here, with nonpole terms providing at most small-angle corrections. More generally, the Debye-CAM formulation makes it possible to relate prominent angular structures of the scattering cross section to distinct physical mechanisms by separating exactly the surface-associated and interior-associated pole content. In particular, for neutron star-like compactness at high frequency, we show that the  rainbow-like enhancement already appears at the first interior-transmission order and is primarily associated with the Regge-Debye pole family linked to the interior regularity condition.

The paper is organized as follows. In Sec.~\ref{sec:waves} we set up the framework for wave scattering by a horizonless compact-object spacetime, specifying the spacetime model, the effective potential, and the boundary conditions defining the $S$-matrix. In Sec.~\ref{sec:debye} we derive the Debye-series expansion of the $S$-matrix and discuss its interpretation: we introduce the exterior ingoing/upgoing basis and the local interior surface basis together with the associated surface connection coefficients (Sec.~\ref{sec:debye_1}), define the regular interior solution and the internal phase factor (Sec.~\ref{sec:debye_2}), and obtain the Debye-series decomposition (Sec.~\ref{sec:debye_3}). In Sec.~\ref{sec:spectrum} we compute and classify the Regge-Debye pole spectrum; the numerical method is summarized in Sec.~\ref{subsec:numerics}, and the resulting spectra (including the identification of three pole branches) are presented in Sec.~\ref{sec: Results RDP spectrum}. In Sec.~\ref{sec:results_debye} we reconstruct the differential scattering cross section from the Debye partial-wave contributions and compare with direct partial-wave calculations. After recalling the standard partial-wave formulas (Sec.~\ref{sec:results_debye_1}) and introducing the Debye decomposition of the amplitude (Sec.~\ref{sec:results_debye_2}), we describe the numerical implementation (Sec.~\ref{sec:results_debye_3}) and present representative results (Sec.~\ref{subsec:debye_scatt_results}). In Sec.~\ref{sec:CAM_debye} we develop CAM representations of the Debye amplitudes: we first present the modified Sommerfeld-Watson transform appropriate to the branch-cut structure of the present problem (Sec.~\ref{sec:CAM_debye_1}), then derive the CAM representations (Sec.~\ref{subsec:cam_debye_rep}), and finally present the numerical implementation and CAM reconstructions together with pole/cut/background decompositions (Secs.~\ref{sec:CAM_debye_3} and~\ref{sec:CAM_debye_4}). We conclude in Sec.~\ref{sec:conclusion} with a summary and outlook.

Throughout this article, we adopt natural units such that $\hbar = c = G = 1$. We also assume a harmonic time dependence of the form $e^{-i \omega t}$ for the perturbative field.

	%=========================================================
	\section{Scalar waves on a compact-object spacetime}
	\label{sec:waves}
	
	In this section we fix notation and summarize the basic framework for the scattering of a massless scalar wave by a horizonless compact body. The setup follows our earlier CAM study of compact-object scattering~\cite{OuldElHadj:2019kji}.
	
	We consider a static, spherically symmetric spacetime describing a compact object of areal radius $R$ with a regular center at $r=0$. In Schwarzschild-like coordinates $\{t,r,\theta,\phi\}$ the line element reads
	\begin{equation}
		\dd s^2=-F(r)\,\dd t^2+H(r)^{-1}\,\dd r^2+r^2\,\dd\sigma^2,
		\label{eq:metric}
	\end{equation}
	where $\dd\sigma^2=\dd\theta^2+\sin^2\theta\,\dd\phi^2$ is the metric on the unit $2$-sphere $S^2$. The exterior region $r>R$ is vacuum and is therefore Schwarzschild by Birkhoff's theorem~\cite{VojeJohansen:2005nd},
	\begin{equation}
		F(r)=H(r)=1-\frac{2M}{r},
		\label{eq:schw_ext}
	\end{equation}
	with $M$ the total mass of the object. The interior functions $F(r)$ and $H(r)$ are specified by a stellar model. In this work we adopt, as in Ref.~\cite{OuldElHadj:2019kji}, the incompressible perfect fluid (constant density) model~\cite{Shapiro1983}, for which
	\begin{subequations}
		\label{eq:EoS}
		\begin{align}
				\rho&=\frac{M}{\frac{4}{3}\pi R^{3}},\\
				p(r)&=\rho\,\frac{\beta(R)-\beta(r)}{\beta(r)-3\beta(R)},\\
				\beta(x)&=\sqrt{3-8\pi\rho\,x^{2}},
			\end{align}
	\end{subequations}
	and the interior metric functions are
	\begin{subequations}\label{Interior_Solution}
		\begin{align}\label{Interior_Solution_f}
				F(r) &= \frac{1}{4}\left(1-\frac{2 M r^2}{R^3}\right)+\frac{9}{4}\left(1-\frac{2M}{R}\right) \nonumber \\
				&\quad -\frac{3}{2} \sqrt{\left(1-\frac{2M}{R}\right)\left(1-\frac{2 M r^2}{R^3}\right)}, \\
				H(r) &= 1-\frac{2 M r^2}{R^3}.
				\label{Interior_Solution_h}
			\end{align}
	\end{subequations}
	
	We study a massless scalar field $\Phi$ satisfying the Klein-Gordon equation
	\begin{equation}
		\Box \Phi = \frac{1}{\sqrt{-g}}\partial_\mu\!\left(\sqrt{-g}\,g^{\mu\nu}\partial_\nu\Phi\right)=0.
		\label{eq:KG}
	\end{equation}
	With the separation of variables
	\begin{equation}
		\Phi(t,r,\theta,\phi)=\frac{1}{r}\sum_{\ell=0}^{\infty}\sum_{m=-\ell}^{\ell}
		\phi_{\omega\ell}(r)\,Y_{\ell m}(\theta,\phi)\,e^{-i\omega t},
		\label{eq:sep}
	\end{equation}
	one obtains the radial equation
	\begin{equation}
		\frac{\dd^2\phi_{\omega\ell}}{\dd r_\ast^{\,2}}+\bigl[\omega^2-V_\ell(r)\bigr]\phi_{\omega\ell}=0,
		\label{eq:radial}
	\end{equation}
	where the tortoise coordinate $r_\ast$ is defined by
	\begin{equation}
		\frac{\dd r_\ast}{\dd r}=\frac{1}{\sqrt{F(r)H(r)}},
		\label{eq:tortoise}
	\end{equation}
	with the additive constant chosen so that $r_\ast(r)$ is continuous at $r=R$.
	
	For a scalar field, the effective potential can be written as
	\begin{equation}
		V_{\ell}(r)=F(r)\left[\frac{\ell(\ell+1)}{r^{2}}
		+\frac{H(r)}{2r}\left(\frac{F'(r)}{F(r)}+\frac{H'(r)}{H(r)}\right)\right],
		\label{eq:V_general_scalar}
	\end{equation}
	where a prime denotes differentiation with respect to $r$. In the Schwarzschild exterior, Eq.~\eqref{eq:V_general_scalar} reduces to the familiar Regge-Wheeler-type potential
	\begin{equation}
		V_{\ell}^{\rm (ext)}(r)=\left(1-\frac{2M}{r}\right)\left[\frac{\ell(\ell+1)}{r^{2}}+\frac{2M}{r^{3}}\right].
		\label{eq:Vext}
	\end{equation}
	
	Because the matter content changes at $r=R$, $V_{\ell}(r)$ typically exhibits a finite jump,
	\begin{equation}
		\Delta V_{\ell}\equiv \lim_{\epsilon\to0}\!\left[V_{\ell}(R+\epsilon)-V_{\ell}(R-\epsilon)\right]\neq0,
		\label{eq:jump_def}
	\end{equation}
	and for the incompressible model this jump is positive in the scalar case,
	\begin{equation}
		\Delta V_{\ell}=+\frac{3M F(R)}{R^{3}}.
	\end{equation}
	
	The physical scattering solution is selected by regularity at the origin. From Eq.~\eqref{eq:radial} one finds, as $r\to 0$,
	\begin{equation}
		\phi_{\omega\ell}(r)\propto r^{\ell+1},
		\label{eq:bc_center}
	\end{equation}
	which fixes the interior solution up to normalization. At the surface $r=R$ we impose continuity of the field and of its first $r_\ast$ derivative ($C^1$ matching in $r_\ast$). In the exterior region, as $r_\ast\to+\infty$, the field admits the standard ingoing/outgoing decomposition
	\begin{equation}
		\phi_{\omega\ell}(r)\underset{r_\ast\to+\infty}{\sim}
		A^{(-)}_\ell(\omega)\,e^{-i\omega r_\ast}
		+A^{(+)}_\ell(\omega)\,e^{+i\omega r_\ast},
		\label{eq:asympt}
	\end{equation}
	where $A^{(\pm)}_\ell(\omega)$ are complex amplitudes. They define the $S$-matrix element
	\begin{equation}
		S_\ell(\omega)=e^{i(\ell+1)\pi}\,\frac{A^{(+)}_\ell(\omega)}{A^{(-)}_\ell(\omega)}.
		\label{eq:Smatrix}
	\end{equation}
	For integer $\ell$ and real $\omega$, $S_\ell(\omega)$ determines the phase shifts and the elastic scattering amplitude through the standard partial-wave series~\cite{Futterman:1988ni}. In the CAM approach one analytically continues $\ell\mapsto\lambda-\tfrac12$ and studies the pole structure of $S_{\lambda-1/2}(\omega)$ in the complex $\lambda$-plane at fixed real $\omega$. As shown in Ref.~\cite{OuldElHadj:2019kji}, compact objects may support, besides the familiar surface-wave branch, an additional broad-resonance branch associated with the surface behavior of the effective potential. In the ultracompact regime ($R<3M$), the exterior barrier near the light ring can combine with the stellar surface to form a potential cavity in $r_\ast$, leading to long-lived (narrow) resonances~\cite{OuldElHadj:2019kji}.
	
	Building on this framework, we now introduce an exact Debye-series expansion of the ratio $A^{(+)}/A^{(-)}$ tailored to compact bodies. This yields a decomposition of the $S$-matrix element into contributions associated with direct surface reflection and with successive interior traversals. In turn, it allows us to define and interpret the corresponding Regge-Debye pole families and to reconstruct each Debye order within the CAM formalism.

    %=========================================================
    \section{The Debye expansion}
    \label{sec:debye}
    
    In this section we derive an exact Debye-series decomposition of the $S$-matrix element~\eqref{eq:Smatrix}. The resulting expansion separates the scattering process into contributions associated with direct surface reflection and with interior propagation, in which waves enter the compact body, undergo one or more interior traversals with partial reflection at the surface, and eventually reemerge to infinity. More precisely, this is achieved by rewriting the ratio $A^{(+)}/A^{(-)}$ in terms of effective surface reflection and transmission factors together with an internal phase factor. Expanding the resulting Fabry-P\'erot-like denominator as a geometric series then yields the Debye series. This construction parallels the classical Debye expansion in wave optics~\cite{Nussenzveig:2006}, while adapted here to compact-body spacetimes.
    
    %---------------------------------------------------------
    \subsection{Exterior ingoing/upgoing basis and interior surface basis}
    \label{sec:debye_1}
      
    In the exterior Schwarzschild region ($r>R$) we use two independent solutions normalized by their asymptotic behavior at spatial infinity,
    \begin{equation}
    	\left\{
    	\begin{aligned}
    		f^{\rm in}_{\omega\ell}(r) &\sim e^{-i\omega r_\ast},\\
    		f^{\rm up}_{\omega\ell}(r) &\sim e^{+i\omega r_\ast},
    	\end{aligned}
    	\right.
    	\qquad (r_\ast\to +\infty),
    	\label{eq:Jost_asympt}
    \end{equation}
    and with Wronskian normalization
    \begin{equation}
    	\W{f^{\rm in}_{\omega\ell}}{f^{\rm up}_{\omega\ell}} = 2 i\omega .
    	\label{eq:WJost}
    \end{equation}
    Throughout we use $\W{f}{g}=f\,\partial_{r_\ast}g-g\,\partial_{r_\ast}f$.
    
    Inside the body ($r<R$) we introduce a \emph{local} interior basis $\{u^{\rm in}_{\omega\ell},\,u^{\rm out}_{\omega\ell}\}$ at the surface by prescribing their values and $r_\ast$ derivatives at $r_\ast=R_\ast$
    \begin{equation}
    	\left\{
    	\begin{aligned}
    		u^{\rm in}_{\omega\ell}(R_\ast) &= 1,  & \quad \left.\partial_{r_\ast}u^{\rm in}_{\omega\ell}\right|_{R_\ast} &= - i\,k_{\rm int},\\
    		u^{\rm out}_{\omega\ell}(R_\ast) &= 1, & \quad \left.\partial_{r_\ast}u^{\rm out}_{\omega\ell}\right|_{R_\ast} &= + i\,k_{\rm int},
    	\end{aligned}
    	\right.
    	\label{eq:uin_uout_surface}
    \end{equation}
    where
    \begin{equation}
    	k_{\rm int}=k_{\rm int}(\omega,\ell)=\sqrt{\omega^2-V_\ell(R^-)} .
    	\label{eq:kint_def}
    \end{equation}
    Here $V_\ell(R^-)$ denotes the interior limit of the effective potential at the surface, and Eq.~\eqref{eq:uin_uout_surface} immediately gives
    \begin{equation}
    	\W{u^{\rm in}_{\omega\ell}}{u^{\rm out}_{\omega\ell}}=2 i k_{\rm int}.
    	\label{eq:Wuin_uout}
    \end{equation}
	
	We emphasize that the labels ``in/out'' for the interior basis and ``in/up'' for the exterior basis refer to distinct local propagation conventions. In particular, the label ``in'' is used in both regions but corresponds to different physical directions: it denotes inward propagation from spatial infinity in the exterior region, whereas it describes propagation toward the center in the interior region. This distinction reflects the standard definitions adopted in each region and must be kept in mind when interpreting the matching conditions at the surface.

    At the surface $r=R$ (or equivalently, $r_\ast=R_\ast$), the interior and exterior radial solutions are matched by requiring continuity of the field and of its first $r_\ast$ derivative. This yields linear connection relations between the exterior basis $\{f^{\rm in}_{\omega\ell},f^{\rm up}_{\omega\ell}\}$ and the local interior basis $\{u^{\rm in}_{\omega\ell},u^{\rm out}_{\omega\ell}\}$. In particular, expanding the interior ingoing mode in the exterior basis,
    \begin{equation}
    	u^{\rm in}_{\omega\ell}(r_\ast)
    	=\alpha^{\rm in}_\ell(\omega)\,f^{\rm in}_{\omega\ell}(r_\ast)
    	+\beta^{\rm out}_\ell(\omega)\,f^{\rm up}_{\omega\ell}(r_\ast),
    	\label{eq:uin_expand}
    \end{equation}
    and imposing the matching conditions at $r_\ast=R_\ast$ determines $\alpha^{\rm in}_\ell(\omega)$ and $\beta^{\rm out}_\ell(\omega)$. Solving the resulting $2\times2$ linear system gives the Wronskian expressions
    \begin{subequations}\label{eq:alpha_beta}
    	\begin{align}
    		\alpha^{\rm in}_\ell(\omega)&=\frac{\W{u^{\rm in}_{\omega\ell}}{f^{\rm up}_{\omega\ell}}}{\W{f^{\rm in}_{\omega\ell}}{f^{\rm up}_{\omega\ell}}}\Bigg|_{r_\ast=R_\ast},
    		\label{eq:alpha_beta_a}\\
    		\beta^{\rm out}_\ell(\omega)&=\frac{\W{f^{\rm in}_{\omega\ell}}{u^{\rm in}_{\omega\ell}}}{\W{f^{\rm in}_{\omega\ell}}{f^{\rm up}_{\omega\ell}}}\Bigg|_{r_\ast=R_\ast}.
    		\label{eq:alpha_beta_b}
    	\end{align}
    \end{subequations}
    
    Conversely, expanding the exterior upgoing solution in the interior basis,
    \begin{equation}
    	f^{\rm up}_{\omega\ell}(r_\ast)=\gamma^{\rm out}_\ell(\omega)\,u^{\rm out}_{\omega\ell}(r_\ast)+\delta^{\rm in}_\ell(\omega)\,u^{\rm in}_{\omega\ell}(r_\ast),
    	\label{eq:fup_expand}
    \end{equation}
    the matching conditions at $r_\ast=R_\ast$ determine $\gamma^{\rm out}_\ell(\omega)$ and $\delta^{\rm in}_\ell(\omega)$, which can likewise be expressed as
    \begin{subequations}\label{eq:gamma_delta}
    	\begin{align}
    		\gamma^{\rm out}_\ell(\omega)&=\frac{\W{f^{\rm up}_{\omega\ell}}{u^{\rm in}_{\omega\ell}}}{\W{u^{\rm out}_{\omega\ell}}{u^{\rm in}_{\omega\ell}}}\Bigg|_{r_\ast=R_\ast},
    		\label{eq:gamma_delta_a}\\
    		\delta^{\rm in}_\ell(\omega)&=\frac{\W{u^{\rm out}_{\omega\ell}}{f^{\rm up}_{\omega\ell}}}{\W{u^{\rm out}_{\omega\ell}}{u^{\rm in}_{\omega\ell}}}\Bigg|_{r_\ast=R_\ast}.
    		\label{eq:gamma_delta_b}
    	\end{align}
    \end{subequations}
    
    It is convenient to collect Eq.~\eqref{eq:uin_expand} together with the analogous expansion of $u^{\rm out}_{\omega\ell}$ in the exterior basis into the matrix form
    \begin{equation}
    	\binom{u^{\rm in}_{\omega\ell}}{u^{\rm out}_{\omega\ell}}
    	=  \mathbf{M}
    	\binom{f^{\rm in}_{\omega\ell}}{f^{\rm up}_{\omega\ell}},
    	\label{eq:conn_matrix}
    \end{equation}
    where
    \begin{equation}
    	\mathbf{M}
    	=
    	\begin{pmatrix}
    		\alpha^{\rm in}_\ell(\omega) & \beta^{\rm out}_\ell(\omega)\\
    		\alpha^{\rm out}_\ell(\omega) & \beta^{\rm in}_\ell(\omega)
    	\end{pmatrix}.
    	\label{eq:conn_matrix_bis}
    \end{equation}
    In Eqs.~\eqref{eq:conn_matrix}--\eqref{eq:conn_matrix_bis}, $\alpha^{\rm out}_\ell(\omega)$ and $\beta^{\rm in}_\ell(\omega)$ are defined by expanding $u^{\rm out}_{\omega\ell}$ in the same way as \eqref{eq:uin_expand}. The determinant of the connection matrix $\mathbf{M}$ is fixed by the surface Wronskians,
    \begin{equation}
    	\det M=\frac{\W{u^{\rm in}_{\omega\ell}}{u^{\rm out}_{\omega\ell}}}{\W{f^{\rm in}_{\omega\ell}}{f^{\rm up}_{\omega\ell}}}
    	=\frac{k_{\rm int}}{\omega}.
    	\label{eq:detM}
    \end{equation}
    Inverting \eqref{eq:conn_matrix} and comparing with Eq.~\eqref{eq:fup_expand} then yields, in particular,
    \begin{equation}
    	\gamma^{\rm out}_\ell(\omega)=\frac{\alpha^{\rm in}_\ell(\omega)}{\det M}
    	=\frac{\omega}{k_{\rm int}}\,\alpha^{\rm in}_\ell(\omega).
    	\label{eq:gamma_alpha_relation}
    \end{equation}
    
    \subsection{Regular interior solution and the internal phase factor}
    \label{sec:debye_2}
    
    Let $\phi^{\rm reg}_{\omega\ell}(r)$ denote the solution of the radial equation that is regular at the center, integrated up to the surface $r=R$. Since $\{u^{\rm in}_{\omega\ell},u^{\rm out}_{\omega\ell}\}$ form a basis in the interior region, $\phi^{\rm reg}_{\omega\ell}$ can be expanded as
    \begin{equation}
    	\phi^{\rm reg}_{\omega\ell}(r)=c^{\rm in}_\ell(\omega)\,u^{\rm in}_{\omega\ell}(r)+c^{\rm out}_\ell(\omega)\,u^{\rm out}_{\omega\ell}(r),
    	\label{eq:phireg_expand_u}
    \end{equation}
    with coefficients given by the Wronskian identities
    \begin{subequations}\label{eq:cin_cout}
    	\begin{align}
    		c^{\rm in}_\ell(\omega)&=\frac{\W{\phi^{\rm reg}_{\omega\ell}}{u^{\rm out}_{\omega\ell}}}{\W{u^{\rm in}_{\omega\ell}}{u^{\rm out}_{\omega\ell}}}\Bigg|_{r_\ast=R_\ast},
    		\label{eq:cin_cout_a}\\
    		c^{\rm out}_\ell(\omega)&=\frac{\W{u^{\rm in}_{\omega\ell}}{\phi^{\rm reg}_{\omega\ell}}}{\W{u^{\rm in}_{\omega\ell}}{u^{\rm out}_{\omega\ell}}}\Bigg|_{r_\ast=R_\ast}.
    		\label{eq:cin_cout_b}
    	\end{align}
    \end{subequations}
    We then define the internal phase factor
    \begin{equation}
    	\xi_\ell(\omega)\equiv \frac{c^{\rm out}_\ell(\omega)}{c^{\rm in}_\ell(\omega)}
    	=\frac{\W{u^{\rm in}_{\omega\ell}}{\phi^{\rm reg}_{\omega\ell}}}{\W{\phi^{\rm reg}_{\omega\ell}}{u^{\rm out}_{\omega\ell}}}\Bigg|_{r_\ast=R_\ast},
    	\label{eq:xi_def}
    \end{equation}
    which corresponds to propagation from the surface to the center and back.
    
    %---------------------------------------------------------
    \subsection{Debye series}
    \label{sec:debye_3}
    
    Starting from the interior expansion \eqref{eq:phireg_expand_u}, we use the surface connection relations \eqref{eq:uin_expand} and \eqref{eq:fup_expand} to rewrite the regular solution $\phi^{\rm reg}_{\omega\ell}$ in the exterior basis $\{f^{\rm in}_{\omega\ell},f^{\rm up}_{\omega\ell}\}$:
    \begin{equation}
    	\begin{split}
    		\phi^{\rm reg}_{\omega\ell}(r)
    		= &\alpha^{\rm in}_\ell\,c^{\rm in}_\ell\Bigl(1-\frac{\delta^{\rm in}_\ell}{\gamma^{\rm out}_\ell}\,\xi_\ell\Bigr)\,f^{\rm in}_{\omega\ell}(r) \\
    		& + \Big(\beta^{\rm out}_\ell\,c^{\rm in}_\ell\Bigl(1-\frac{\delta^{\rm in}_\ell}{\gamma^{\rm out}_\ell}\,\xi_\ell\Bigr)
    		+ \frac{1}{\gamma^{\rm out}_\ell}\,c^{\rm out}_\ell\Big)\,f^{\rm up}_{\omega\ell}(r).
    	\end{split}
    	\label{eq:phireg_expand_ext_repeat_bis}
    \end{equation}
    
    Comparing with the far-field decomposition
    \begin{equation}
    	\phi^{\rm reg}_{\omega\ell}(r)
    	= A^{(-)}_\ell(\omega)\,f^{\rm in}_{\omega\ell}(r) + A^{(+)}_\ell(\omega)\,f^{\rm up}_{\omega\ell}(r),
    	\label{eq:phireg_expand_ext_repeat}
    \end{equation}
    we obtain [cf.~\eqref{eq:gamma_alpha_relation}]
    \begin{equation}
    	\begin{aligned}
    		A^{(-)}_\ell(\omega)
    		&= \alpha^{\rm in}_\ell\,c^{\rm in}_\ell
    		- \frac{k_{\rm int}}{\omega}\,c^{\rm out}_\ell\,\delta^{\rm in}_\ell  \\
    		&= \alpha^{\rm in}_\ell\,c^{\rm in}_\ell\Bigl(1-\frac{\delta^{\rm in}_\ell}{\gamma^{\rm out}_\ell}\,\xi_\ell\Bigr),
    	\end{aligned}
    	\label{eq:Aminus_Debye}
    \end{equation}
    and
    \begin{equation}
    	A^{(+)}_\ell(\omega)=\beta^{\rm out}_\ell\,c^{\rm in}_\ell\Bigl(1-\frac{\delta^{\rm in}_\ell}{\gamma^{\rm out}_\ell}\,\xi_\ell\Bigr)
    	+\frac{1}{\gamma^{\rm out}_\ell}\,c^{\rm out}_\ell.
    	\label{eq:Aplus_Debye}
    \end{equation}

	Combining expressions \eqref{eq:Aminus_Debye} and  \eqref{eq:Aplus_Debye}, we obtain an exact cavity-like representation of the scattering process. It is then natural to introduce effective reflection and transmission coefficients at the surface. Following the standard convention in the theory of Debye series~\cite{Nussenzveig:2006}, we denote the interior region ($r<R$) by the index $1$ and the exterior region ($r>R$) by $2$. In this notation, the ratio $A^{(+)}/A^{(-)}$ can be written as
	\begin{equation}
		\frac{A^{(+)}_\ell(\omega)}{A^{(-)}_\ell(\omega)}
		=
		R_{22,\ell}
		+\frac{T_{12,\ell}\,T_{21,\ell}\,\xi_\ell}{1-R_{11,\ell}\,\xi_\ell},
		\label{eq:Rtot_cavity}
	\end{equation}
	where $R_{ii,\ell}$ denotes reflection back into region $i$, while $T_{ij,\ell}$ denotes transmission from region $j$ to region $i$. The effective surface coefficients are defined by
	\begin{subequations}\label{eq:Debye_coeffs_def}
		\begin{align}
			R_{22,\ell}(\omega)&\equiv \frac{\beta^{\rm out}_\ell(\omega)}{\alpha^{\rm in}_\ell(\omega)},
			\label{eq:Debye_coeffs_def_a}\\
			T_{21,\ell}(\omega)&\equiv \frac{1}{\alpha^{\rm in}_\ell(\omega)},
			\label{eq:Debye_coeffs_def_b}\\
			T_{12,\ell}(\omega)&\equiv \frac{1}{\gamma^{\rm out}_\ell(\omega)},
			\label{eq:Debye_coeffs_def_c}\\
			R_{11,\ell}(\omega)&\equiv \frac{\delta^{\rm in}_\ell(\omega)}{\gamma^{\rm out}_\ell(\omega)}.
			\label{eq:Debye_coeffs_def_d}
		\end{align}
	\end{subequations}
	As in the semiclassical theory of the rainbow, $R_{22,\ell}$ and $T_{21,\ell}$ describe reflection back to infinity and transmission into the interior for an externally incident wave, while $R_{11,\ell}$ and $T_{12,\ell}$ describe reflection back into the interior and transmission to infinity for an internally incident wave.

    The Debye series follows by expanding the inverse of the denominator in \eqref{eq:Rtot_cavity} as a geometric series,
    \begin{equation}
    	\frac{A^{(+)}_\ell(\omega)}{A^{(-)}_\ell(\omega)}
    	=
    	R_{22,\ell}
    	+\sum_{p=1}^{\infty} T_{12,\ell}T_{21,\ell}\,R_{11,\ell}^{\,p-1}\,\xi_\ell^{\,p}.
    	\label{eq:Rtot_Debye}
    \end{equation}
    This yields the corresponding Debye decomposition of the $S$-matrix element~\eqref{eq:Smatrix},
    \begin{equation}
    	S_\ell(\omega)=S^{(0)}_\ell(\omega)+\sum_{p=1}^{\infty}S^{(p)}_\ell(\omega),
    	\label{eq:S_Debye_sum}
    \end{equation}
    with
    \begin{subequations}\label{eq:S_Debye_terms}
    	\begin{align}
    		S^{(0)}_\ell(\omega)&=e^{i(\ell+1)\pi}\,R_{22,\ell},
    		\label{eq:S_Debye_terms_a}\\
    		S^{(p)}_\ell(\omega)&=e^{i(\ell+1)\pi}\,T_{12,\ell}T_{21,\ell}\,R_{11,\ell}^{\,p-1}\,\xi_\ell^{\,p}.
    		\label{eq:S_Debye_terms_b}
    	\end{align}
    \end{subequations}
    
    This expansion admits a transparent physical interpretation~\cite{Nussenzveig:2006}. The term $S^{(0)}_\ell$ corresponds to direct reflection at the surface, encoded by the exterior reflection coefficient $R_{22,\ell}$. For $p\ge1$, the contribution $S^{(p)}_\ell$ describes a wave that is transmitted into the interior (factor $T_{21,\ell}$), undergoes $p-1$ reflections at the surface from the inside (factor $R_{11,\ell}^{\,p-1}$), accumulates the internal phase $\xi_\ell$ during each surface-center-surface propagation (factor $\xi_\ell^{\,p}$), and finally escapes to infinity through transmission across the surface (factor $T_{12,\ell}$). Regularity at $r=0$ effectively enforces total reflection at the center and closes the interior propagation cycle.
    
    For the CAM analysis, Eqs.~\eqref{eq:S_Debye_terms_a} and \eqref{eq:S_Debye_terms_b} are most conveniently rewritten in terms of the surface connection coefficients $\alpha^{\rm in}_\ell,\beta^{\rm out}_\ell,\gamma^{\rm out}_\ell,\delta^{\rm in}_\ell$ together with the interior phase factor $\xi_\ell$. Using \eqref{eq:Debye_coeffs_def} in \eqref{eq:S_Debye_terms} and replacing $\gamma^{\rm out}_\ell$ by \eqref{eq:gamma_alpha_relation}, we obtain
    \begin{subequations}\label{eq:S_Debye_terms_coeffs}
    	\begin{align}
    		S^{(0)}_\ell(\omega)&=e^{i(\ell+1)\pi}\,\frac{\beta^{\rm out}_\ell(\omega)}{\alpha^{\rm in}_\ell(\omega)},
    		\label{eq:S_Debye_terms_coeffs_a}\\
    		S^{(p)}_\ell(\omega)&=e^{i(\ell+1)\pi}\,
    		\left(\frac{k_{\rm int}(\omega,\ell)}{\omega}\right)^{p}\,
    		\frac{\xi_\ell(\omega)^{\,p}\,\Bigl(\delta^{\rm in}_\ell(\omega)\Bigr)^{p-1}}{\Bigl(\alpha^{\rm in}_\ell(\omega)\Bigr)^{p+1}}.
    		\label{eq:S_Debye_terms_coeffs_b}
    	\end{align}
    \end{subequations}
    
    In this form, the analytic continuation to complex $\lambda$ is straightforward. The Regge-Debye pole structure is then determined by thee zeros $\alpha^{\rm in}_{\lambda-1/2}$ and, for $p\ge1$, by the additional singular structure entering through $\xi_{\lambda-1/2}=c^{\rm out}_{\lambda-1/2}/c^{\rm in}_{\lambda-1/2}$.

%=========================================================
\section{The Regge-Debye pole spectrum}
\label{sec:spectrum}

\textit{The Regge-Debye pole spectrum} is the set of complex angular momenta $\lambda_n(\omega)\equiv \ell_n(\omega)+1/2$ at which the analytically continued Debye $S$-matrix elements \eqref{eq:S_Debye_sum} [see also Eqs.~\eqref{eq:S_Debye_terms_coeffs_a} and \eqref{eq:S_Debye_terms_coeffs_b}] have simple poles for fixed real frequency $\omega$ (i.e.\ $\omega\in\mathbb{R}$ and $\lambda_n(\omega)\in\mathbb{C}$). Here $n=1,2,\dots$ labels the discrete sequence of poles in the complex $\lambda$-plane. In the compact-body problem considered here, the Regge-Debye spectrum is governed by two distinct mechanisms: a surface contribution associated with the zeros of the surface coefficient $\alpha^{\rm in}_{\lambda-1/2}(\omega)$, and an interior contribution carried by the internal phase factor $\xi_{\lambda-1/2}(\omega)$, i.e., by the zeros of $c^{\rm in}_{\lambda-1/2}(\omega)$.

%=================================================================
\subsection{Numerical method}
\label{subsec:numerics}

To compute the Regge-Debye pole spectrum associated with the surface coefficient $\alpha^{\rm in}_{\lambda-1/2}(\omega)$, we adapt the numerical procedure developed in Ref.~\cite{OuldElHadj:2019kji} for locating Regge poles from the outgoing condition at infinity. We therefore summarize only the main steps and highlight the modifications required in the present setting.

Following Ref.~\cite{OuldElHadj:2019kji}, we represent an exterior solution of the radial Eq.~\eqref{eq:radial} in the Schwarzschild region by a power series about a point $r=b>R$,
\begin{equation}
	\phi(r)=e^{+i\omega r_\ast(r)}\sum_{n=0}^{\infty} a_n\left(1-\frac{b}{r}\right)^n,
	\label{eq:series_b}
\end{equation}
where the coefficients $a_n$ satisfy a four-term recurrence relation
\begin{equation}
	\alpha_n a_{n+1} + \beta_n a_{n} +\gamma_{n} a_{n-1} +\delta_{n} a_{n-2}  = 0,
	\quad \forall n\ge 2,
	\label{eq:rec4}
\end{equation}
with coefficients
\begin{subequations}
	\begin{eqnarray}\label{Coeffs_3_termes}
		&& \alpha_n = n (n+1)\left(1-\frac{2M}{b}\right), \\
		&& \beta_n  = 2i\omega b\, n + 3\left(\frac{2M}{b}\right)n^2-2n^2, \\
		&& \gamma_n = \left(1-\frac{6M}{b}\right)n(n-1)-\frac{2M}{b}-\ell(\ell+1), \\
		&& \delta_n = \left(\frac{2M}{b}\right)\left(n-1\right)^2 .
	\end{eqnarray}
\end{subequations}

The initial conditions $a_0$ and $a_1$ are fixed by matching the series solution \eqref{eq:series_b} to a numerically integrated exterior solution at $r=b$, i.e.\ by imposing continuity of $\phi$ and $d\phi/dr$ at that point. This yields
\begin{subequations}\label{eq:a0a1}
	\begin{align}
		a_0 &= e^{-i\omega r_\ast(b)}\,\phi(b),
		\label{eq:a0a1_a}\\
		a_1 &= b\,e^{-i\omega r_\ast(b)}
		\left[\frac{d\phi}{dr}\Big|_{r=b}-\frac{i\omega b}{b-2M}\,\phi(b)\right].
		\label{eq:a0a1_b}
	\end{align}
\end{subequations}
Here $\phi(b)$ and $(d\phi/dr)|_{b}$ are obtained from direct numerical integration of the exterior equation from the surface to $r=b$. In the present work, the exterior integration is initialized at $r=R$ with the local ingoing surface conditions
\begin{subequations}\label{eq:surf_ic_alpha}
	\begin{align}
		\phi(R) &= u^{\rm in}(R)=1,
		\label{eq:surf_ic_alpha_a}\\
		\left.\frac{d\phi}{dr}\right|_{r=R} &=
		\left.\frac{du^{\rm in}}{dr}\right|_{r=R}=-i \frac{k_{\rm int}}{F(R)}.
		\label{eq:surf_ic_alpha_b}
	\end{align}
\end{subequations}
and the solution is then propagated outward to the matching point $r=b$.

As in Ref.~\cite{OuldElHadj:2019kji}, we enforce the minimal-solution (purely outgoing) condition for the recurrence \eqref{eq:rec4} via the Hill determinant approach: nontrivial solutions exist only when the corresponding truncated Hill determinant vanishes. The Regge-Debye poles associated with $\alpha^{\rm in}_{\lambda-1/2}(\omega)$ are therefore obtained by fixing $\omega$
and searching for the complex roots $\lambda_n(\omega)$ of the Hill determinant using a Newton iteration in the complex $\lambda$-plane, with the truncation order increased until convergence is achieved.

We next compute the Regge-Debye poles associated with $c^{\rm in}_{\lambda-1/2}$ by solving
\begin{equation}
	\W{\phi^{\rm reg}_{\omega,\lambda-1/2}}{u^{\rm out}_{\omega,\lambda-1/2}}\Big|_{r=R}=0.
	\label{eq:cin_root}
\end{equation}
To this end, we first construct the regular interior solution $\phi^{\rm reg}_{\omega,\lambda-1/2}$ by integrating the radial equation outward from a small radius near the center, $r=\varepsilon$ (typically $\varepsilon\sim 10^{-6}$), up to the surface $r=R$. The integration is initialized at $r=\varepsilon$ with the values $\phi^{\rm reg}_{\omega,\lambda-1/2}(\varepsilon)$ and
$\partial_r\phi^{\rm reg}_{\omega,\lambda-1/2}(\varepsilon)$ determined from a regular Frobenius expansion about $r=0$,
\begin{equation}
	\phi^{\rm reg}_{\omega,\lambda-1/2}(r)=r^{\lambda+1/2}\sum_{n=0}^{N} c_n\,r^n,
	\label{eq:frob_reg}
\end{equation}
where the coefficients $c_n$ are obtained by inserting \eqref{eq:frob_reg} into the radial equation and fixing the normalization by choosing $c_0=1$.

Having obtained $\phi^{\rm reg}_{\omega,\lambda-1/2}(R)$ and its derivative, we evaluate \eqref{eq:cin_root} at $r=R$ and locate its roots in the complex $\lambda$-plane using a Newton iteration.

%=================================================================
\subsection{Results: The Regge-Debye pole spectrum}
\label{sec: Results RDP spectrum}

In this section we present numerical results for the Regge-Debye pole spectrum in two representative configurations: (i) a neutron star-like object with $R/M=6$, and (ii) an UCO with $R/M=2.26$, close to the Buchdahl bound. For each configuration we consider two frequencies: $2M\omega=2$ and $2M\omega=16$ for $R/M=6$, and $2M\omega=2$ and $2M\omega=6$ for $R/M=2.26$.

In what follows we adopt the terminology introduced by Nussenzveig in CAM descriptions of Mie scattering~\cite{Nussenzveig:2006} and already used in Ref.~\cite{OuldElHadj:2019kji}. Although the Regge-Debye poles considered here are distinct from the Regge poles of the full scattering matrix, this choice is motivated by the fact that the Regge-Debye pole spectrum exhibits the same qualitative organization into distinct branches. This difference arises because the Regge poles are determined by the zeros of the full coefficient $A^{(-)}$, whereas the Regge-Debye poles correspond to the zeros of the individual factors $\alpha^{\rm in}$ and $c^{\rm in}$ entering the Debye decomposition. We refer to the poles associated with the zeros of the surface coefficient $\alpha^{\rm in}_{\lambda-1/2}(\omega)$ as the \emph{surface-wave} family. We refer to the poles associated with the zeros of the interior coefficient $c^{\rm in}_{\lambda-1/2}(\omega)$ as \emph{interior resonances}. Depending on the compactness, the interior-resonance sector appears either as a single \emph{broad-resonance} branch or, in the ultracompact regime ($R<3M$), as two sub-branches: broad resonances and an additional \emph{narrow-resonance} branch lying closer to the real axis.

\begin{figure}[!htb]
	\centering
	\includegraphics[scale=0.52]{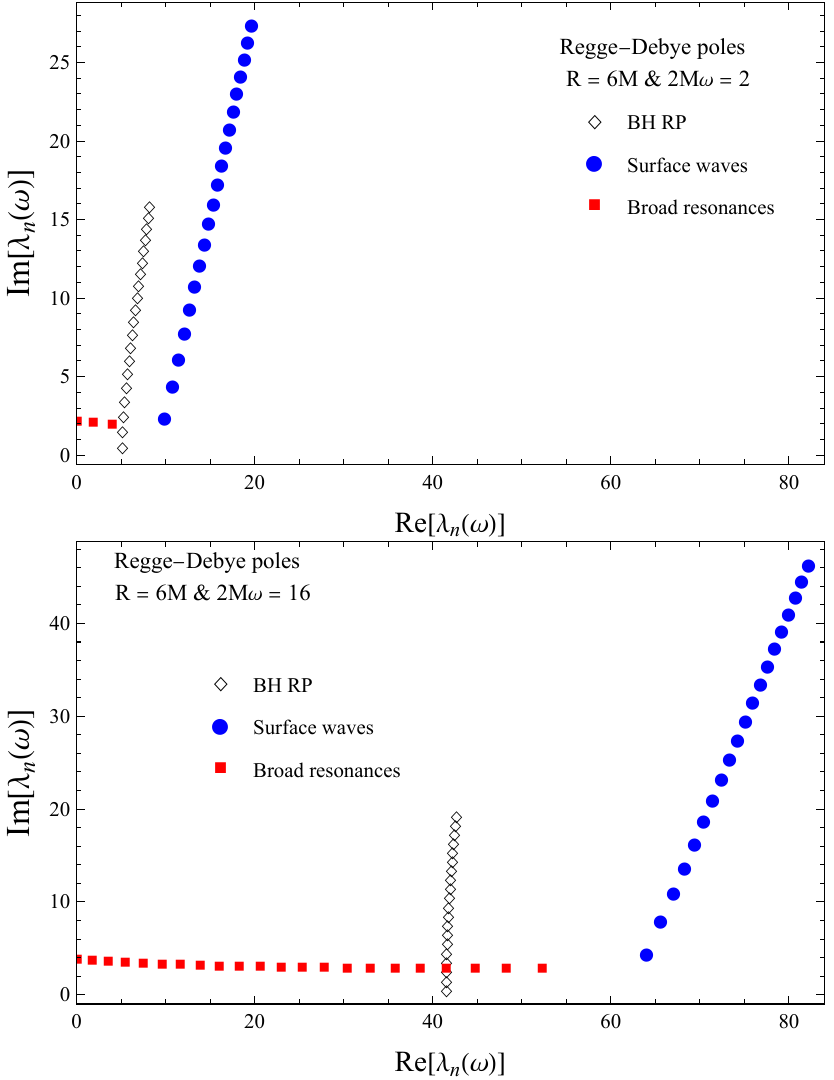}
	\caption{Regge-Debye pole spectrum for the massless scalar field with $R=6M$. The upper panel corresponds to $2M\omega=2$ and the lower panel to $2M\omega=16$. Blue circles denote the surface-wave branch and red squares denote the broad-resonance branch. For comparison, we also show the Regge poles of a Schwarzschild black hole at the same frequency.}
	\label{fig:RDP_R6}
\end{figure}

Figure~\ref{fig:RDP_R6} displays the Regge-Debye spectrum for the neutron star-like configuration $R/M=6$. The surface-wave family (blue circles), obtained from the surface matching condition and associated with the zeros of $\alpha^{\rm in}$, forms an infinite sequence of poles in the first quadrant; as the frequency increases, this sequence shifts toward larger values of $\Re\lambda$. The poles associated with the zeros of $c^{\rm in}$ (red squares) form a distinct broad-resonance family associated with the interior regularity condition. For reference, the Regge poles of a Schwarzschild black hole at the same frequency are shown as black diamonds. The separation between the surface-wave and broad-resonance families is apparent in both panels.
\begin{figure}[htb]
	\centering
	\includegraphics[scale=0.52]{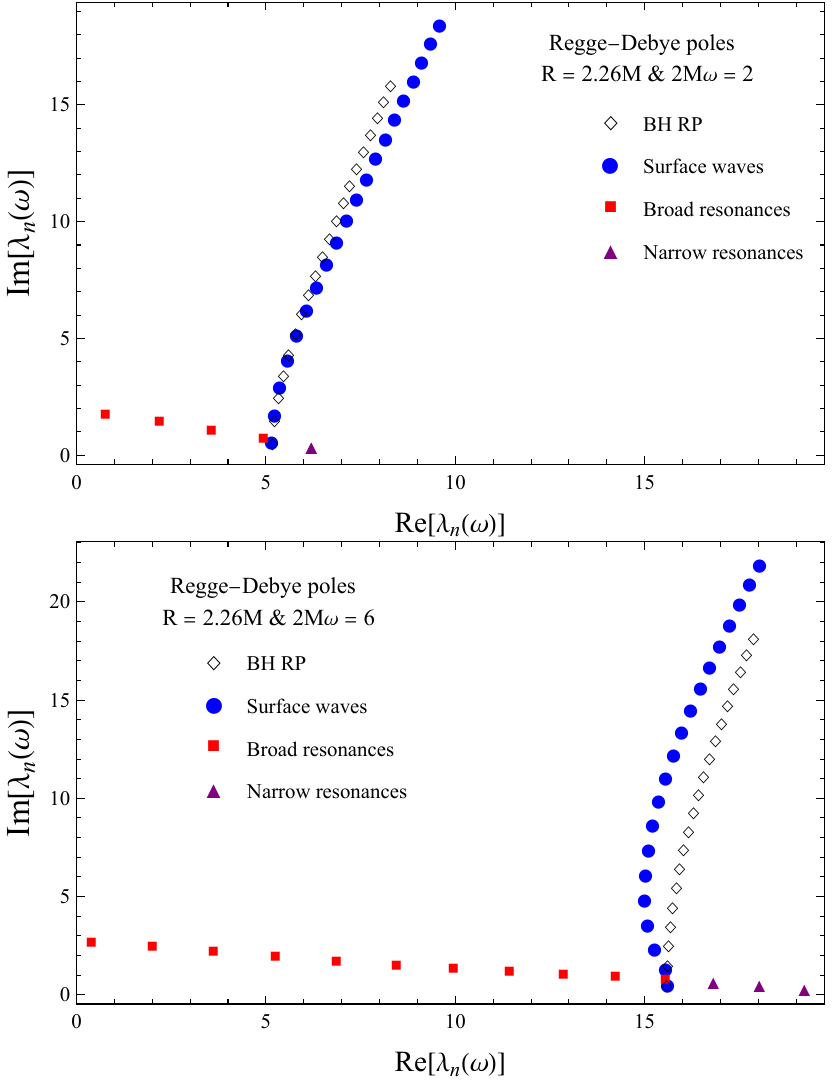}
	\caption{Regge-Debye pole spectrum for the massless scalar field with $R=2.26M$. The upper panel corresponds to $2M\omega=2$ and the lower panel to $2M\omega=6$. Blue circles denote the surface-wave branch, red squares denote the broad-resonance branch, and purple triangles denote the narrow-resonance branch. For comparison, we also show the Regge poles of a Schwarzschild black hole at the same frequency.}
	\label{fig:RDP_R226}
\end{figure}
\begin{figure}[htb]
	\centering
	\includegraphics[scale=0.52]{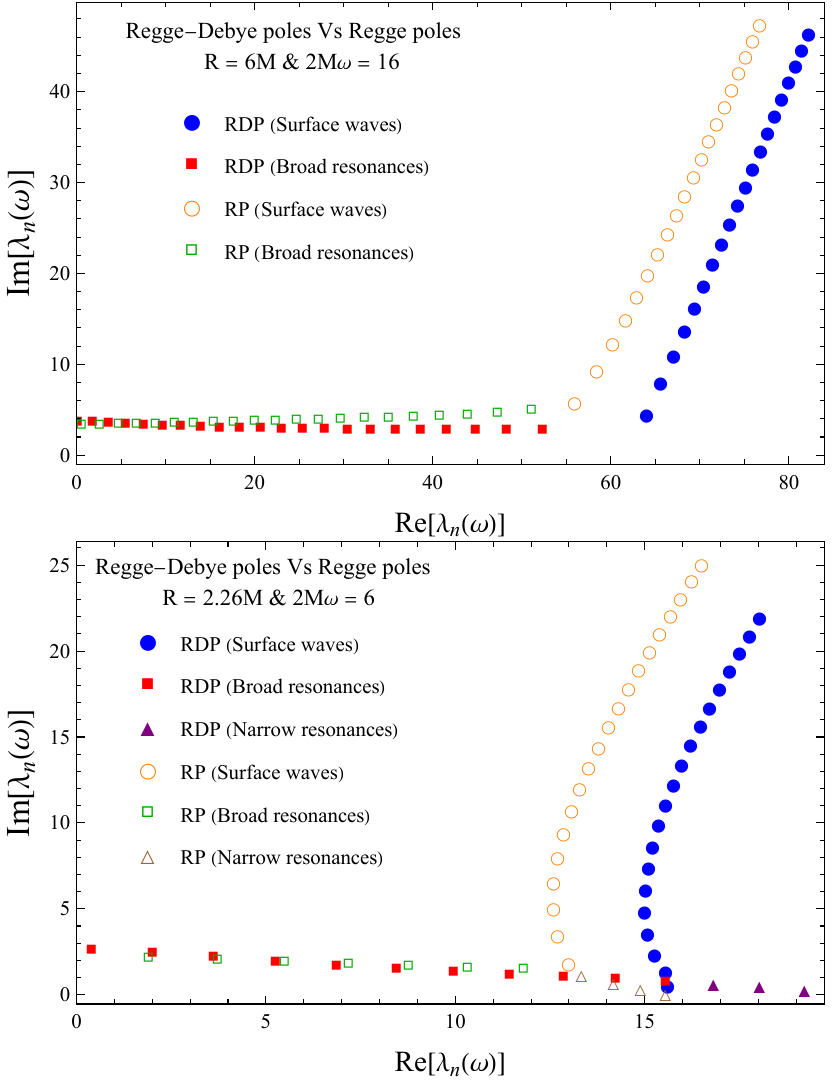}
	\caption{
		Comparison between the Regge-Debye poles and the Regge poles. 
		Top panel: neutron star-like configuration with $R=6M$ at frequency $2M\omega=16$. 
		The Regge-Debye spectrum, obtained from the conditions $\alpha^{\rm in}=0$ and $c^{\rm in}=0$ for the surface-wave and broad-resonance branches, respectively, is compared with the Regge poles obtained from the condition $A^{(-)}=0$ in Ref.~\cite{OuldElHadj:2019kji}. 
		Two families are present in this regime: a surface-wave branch and a broad-resonance branch. 
		Bottom panel: ultracompact configuration with $R=2.26M$ at frequency $2M\omega=6$. 
		In this case, the Regge-Debye surface-wave branch is again generated by the condition $\alpha^{\rm in}=0$, whereas the broad- and narrow-resonance branches are generated by the condition $c^{\rm in}=0$. 
		The Regge poles, by contrast, are obtained in all cases from the single condition $A^{(-)}=0$.}
	\label{RDP_vs_RP_R6_R226}
\end{figure}

Figure~\ref{fig:RDP_R226} shows the corresponding Regge-Debye spectra for the ultracompact case $R/M=2.26$. The surface-wave family (blue circles) persists in the UCO regime and is again associated with the zeros of $\alpha^{\rm in}$. The poles generated by $c^{\rm in}$ split into two subfamilies: a broad-resonance branch (red squares) and a narrow-resonance branch (purple triangles) lying closer to the real axis. The appearance of narrow resonances is a characteristic feature of the UCO regime ($R<3M$), reflecting the presence of long-lived, quasitrapped behavior. As in Fig.~\ref{fig:RDP_R6}, the Schwarzschild black hole Regge poles are included for comparison.

Figure.~\ref{RDP_vs_RP_R6_R226} shows that the Regge-Debye poles and the Regge poles form distinct spectra in the complex angular momentum plane, even though they share the same qualitative branch structure. In particular, the surface-wave and resonance branches identified in the Regge-Debye spectrum have clear counterparts in the spectrum of Regge poles, but their positions in the complex plane are generally shifted. This illustrates the fact that the Debye decomposition isolates the different physical mechanisms contributing to the scattering process (see Sec.~\ref{sec:CAM_debye}), whereas the Regge poles of the full $S$-matrix result from the combined condition $A^{(-)}=0$. In the ultracompact case, this correspondence extends to the narrow-resonance branch, which appears in addition to the surface-wave and broad-resonance branches.

For convenience, the lowest Regge-Debye poles are listed in Table~\ref{tab:R6_RDP_lowest} for $R/M=6$ and in Table~\ref{tab:R226_RDP_lowest} for $R/M=2.26$, grouped by branch (surface wave, broad resonance, and, when present, narrow resonance).
\begin{table*}[t]
	\centering
	\caption{Lowest Regge poles $\lambda_n(\omega)$ for the massless scalar field. The radius of the compact body is $R=6M$.}
	\label{tab:R6_RDP_lowest}
	\small
	\setlength{\tabcolsep}{7pt}
	\renewcommand{\arraystretch}{1.15}
	\begin{tabular}{c c l l}
		\hline\hline
		$2M\omega$ & $n$ & $\lambda^{\mathrm{S\!-\!W}}_n(\omega)$ & $\lambda^{\mathrm{B\!-\!R}}_n(\omega)$ \\
		\hline
		2   & 1  & $9.981874 + 2.349840\,i$  & $0.176661 + 2.087034\,i$ \\
		& 2  & $10.812448 + 4.372277\,i$ & $1.908610 + 2.061304\,i$ \\
		& 3  & $11.515840 + 6.136702\,i$ & $4.098859 + 1.874277\,i$ \\
		& 4  & $12.156244 + 7.756361\,i$ &  \\
		& 5  & $12.755399 + 9.276528\,i$ &  \\
		& 6  & $13.323732 + 10.721831\,i$&  \\
		& 7  & $13.867333 + 12.107802\,i$&  \\
		& 8  & $14.390238 + 13.445115\,i$&  \\
		& 9  & $14.895352 + 14.741525\,i$&  \\
		& 10 & $15.384887 + 16.002891\,i$&  \\
		
		\\[-0.8ex]
		
		16  & 1  & $64.005898 + 4.339915\,i$  & $0.093207 + 3.662691\,i$ \\
		& 2  & $65.672137 + 7.890215\,i$  & $1.827951 + 3.638090\,i$ \\
		& 3  & $67.054318 + 10.913801\,i$ & $3.651179 + 3.538128\,i$ \\
		& 4  & $68.286343 + 13.656500\,i$ & $5.573868 + 3.416709\,i$ \\
		& 5  & $69.420127 + 16.214569\,i$ & $7.576821 + 3.303891\,i$ \\
		& 6  & $70.482872 + 18.638051\,i$ & $9.641521 + 3.207394\,i$ \\
		& 7  & $71.491068 + 20.957046\,i$ & $11.757394 + 3.126461\,i$ \\
		& 8  & $72.455629 + 23.191388\,i$ & $13.919639 + 3.058504\,i$ \\
		& 9  & $73.384222 + 25.355003\,i$ & $16.126860 + 3.001043\,i$ \\
		& 10 & $74.282475 + 27.458141\,i$ & $18.379753 + 2.952090\,i$ \\
		\hline\hline
	\end{tabular}
\end{table*}
\begin{table*}[t]
	\centering
	\caption{Lowest Regge-Debye poles $\lambda_n(\omega)$ for the massless scalar field. The radius of the compact body is $R=2.26M$.}
	\label{tab:R226_RDP_lowest}
	\small
	\setlength{\tabcolsep}{6pt}
	\renewcommand{\arraystretch}{1.15}
	\begin{tabular}{c c l l l}
		\hline\hline
		$2M\omega$ & $n$ &
		$\lambda^{\mathrm{S\!-\!W}}_n(\omega)$ &
		$\lambda^{\mathrm{B\!-\!R}}_n(\omega)$ &
		$\lambda^{\mathrm{N\!-\!R}}_n(\omega)$ \\
		\hline
		2  & 1  & $5.170627 + 0.558427\,i$ & $0.767101 + 1.694005\,i$ & $6.198378 + 0.330026\,i$ \\
		& 2  & $5.229549 + 1.718902\,i$ & $2.198555 + 1.401673\,i$ & \\
		& 3  & $5.379494 + 2.899659\,i$ & $3.580813 + 1.023111\,i$ & \\
		& 4  & $5.589540 + 4.044299\,i$ & $4.931991 + 0.675583\,i$ & \\
		& 5  & $5.831798 + 5.140056\,i$ &  & \\
		& 6  & $6.089716 + 6.189810\,i$ &  & \\
		& 7  & $6.354300 + 7.199744\,i$ &  & \\
		& 8  & $6.620695 + 8.175846\,i$ &  & \\
		& 9  & $6.886261 + 9.123154\,i$ &  & \\
		& 10 & $7.149563 + 10.045752\,i$ &  & \\
		
		\\[-0.8ex]
		
		6  & 1  & $15.619809 + 0.488959\,i$ & $0.394686 + 2.605642\,i$ & $16.819098 + 0.614927\,i$ \\
		& 2  & $15.567624 + 1.300236\,i$ & $2.017115 + 2.423265\,i$ & $18.046027 + 0.467683\,i$ \\
		& 3  & $15.270773 + 2.325872\,i$ & $3.627054 + 2.161910\,i$ & $19.228976 + 0.286427\,i$ \\
		& 4  & $15.089130 + 3.545649\,i$ & $5.255795 + 1.896872\,i$ & \\
		& 5  & $15.017784 + 4.814992\,i$ & $6.869650 + 1.667943\,i$ & \\
		& 6  & $15.028825 + 6.092588\,i$ & $8.443075 + 1.473707\,i$ & \\
		& 7  & $15.102535 + 7.360062\,i$ & $9.966574 + 1.305032\,i$ & \\
		& 8  & $15.224075 + 8.608455\,i$ & $11.437870 + 1.153996\,i$ & \\
		& 9  & $15.382118 + 9.833687\,i$ & $12.857364 + 1.014459\,i$ & \\
		& 10 & $15.568017 + 11.034358\,i$ & $14.226363 + 0.881377\,i$ & \\
		\hline\hline
	\end{tabular}
\end{table*}

%=========================================================
\section{Scattering and Debye-series decomposition}
\label{sec:results_debye}

In this section we recall the standard partial-wave representation of the scattering amplitude and differential cross section for a monochromatic scalar plane wave, and introduce the associated Debye-order decomposition. The term $p=0$ corresponds to direct reflection at the surface, while the terms $p\ge1$ describe waves transmitted into the interior, undergoing $p-1$ internal reflections at the surface, and re-emerging to infinity after $p$ surface-center-surface traversals. This decomposition provides a convenient baseline for the CAM reconstructions developed in Sec.~\ref{sec:CAM_debye}.

%--------------------------------------------------------
\subsection{Partial-wave expansion of the differential scattering cross section}
\label{sec:results_debye_1}

The differential scattering cross section is related to the scattering amplitude by~\cite{Dolan:2017rtj}
\begin{equation}
	\frac{\dd\sigma}{\dd\Omega} = \bigl|f(\omega,\theta)\bigr|^2 .
\end{equation}
The scattering amplitude admits the usual partial-wave expansion
\begin{equation}
	f(\omega,\theta)
	=\frac{1}{2 i \omega}\sum_{\ell=0}^{\infty}(2\ell+1)\bigl[S_\ell(\omega)-1\bigr]P_\ell(\cos\theta),
	\label{eq:pw_series}
\end{equation}
where $P_\ell(\cos\theta)$ are the Legendre polynomials~\cite{AS65}, and the partial-wave $S$-matrix elements $S_\ell(\omega)$ were defined in Eq.~\eqref{eq:Smatrix}.

%%----------------------------------------------------------
%\subsection{Debye expansion of the scattering amplitude}
%\label{sec:results_debye_2}
%
%Substituting the Debye expansion of the $S$-matrix element, Eq.~\eqref{eq:S_Debye_sum}, into the partial-wave series \eqref{eq:pw_series} yields the Debye decomposition
%\begin{equation}
%	f(\omega,\theta)=f^{(0)}(\omega,\theta)+\sum_{p=1}^{\infty}f^{(p)}(\omega,\theta),
%	\label{eq:f_Debye_sum}
%\end{equation}
%where
%\begin{subequations}
%	\begin{align}
%		f^{(0)}(\omega,\theta)
%		&=\frac{1}{2i\omega}\sum_{\ell=0}^{\infty}(2\ell+1)\bigl[S_\ell^{(0)}(\omega)-1\bigr]P_\ell(\cos\theta),
%		\label{eq:f_p_def_a}\\
%		f^{(p)}(\omega,\theta)
%		&=\frac{1}{2i\omega}\sum_{\ell=0}^{\infty}(2\ell+1)\,S_\ell^{(p)}(\omega)\,P_\ell(\cos\theta),
%		\qquad p\ge 1.
%		\label{eq:f_p_def_b}
%	\end{align}
%\end{subequations}
%Here $S_\ell^{(0)}(\omega)$ and $S_\ell^{(p)}(\omega)$ are given by Eqs.~\eqref{eq:S_Debye_terms_coeffs_a} and \eqref{eq:S_Debye_terms_coeffs_b}, respectively. The index $p$ orders the scattering process according to the number of interior traversals between the surface and the center prior to reemergence. The $p=0$ term corresponds to direct reflection at the surface. For $p\ge1$, the $p$th contribution corresponds to waves transmitted into the interior, undergoing $p-1$ internal reflections at the surface, and returning to infinity after $p$ surface-center-surface propagations.

%----------------------------------------------------------
\subsection{Debye expansion of the scattering amplitude}
\label{sec:results_debye_2}

This decomposition separates the scattering amplitude into contributions associated with distinct physical processes. Substituting the Debye expansion of the $S$-matrix element, Eq.~\eqref{eq:S_Debye_sum}, into the partial-wave series \eqref{eq:pw_series} yields the Debye decomposition
\begin{equation}
	f(\omega,\theta)=f^{(0)}(\omega,\theta)+\sum_{p=1}^{\infty}f^{(p)}(\omega,\theta),
	\label{eq:f_Debye_sum}
\end{equation}
where
\begin{subequations}
	\begin{align}
		f^{(0)}(\omega,\theta)
		&=\frac{1}{2i\omega}\sum_{\ell=0}^{\infty}(2\ell+1)\bigl[S_\ell^{(0)}(\omega)-1\bigr]P_\ell(\cos\theta),
		\label{eq:f_p_def_a}\\
		f^{(p)}(\omega,\theta)
		&=\frac{1}{2i\omega}\sum_{\ell=0}^{\infty}(2\ell+1)\,S_\ell^{(p)}(\omega)\,P_\ell(\cos\theta),
		\qquad p\ge 1.
		\label{eq:f_p_def_b}
	\end{align}
\end{subequations}
Here $S_\ell^{(0)}(\omega)$ and $S_\ell^{(p)}(\omega)$ are given by Eqs.~\eqref{eq:S_Debye_terms_coeffs_a} and \eqref{eq:S_Debye_terms_coeffs_b}, respectively. The index $p$ orders the scattering process according to the number of interior traversals between the surface and the center prior to reemergence. The $p=0$ term corresponds to direct reflection at the surface. For $p\ge1$, the $p$th contribution corresponds to waves transmitted into the interior, undergoing $p-1$ internal reflections at the surface, and returning to infinity after $p$ surface-center-surface propagations.

%----------------------------------------------------------
\subsection{Computational methods}
\label{sec:results_debye_3}

The scattering amplitude \eqref{eq:pw_series} is computed using the numerical methods developed in Refs.~\cite{OuldElHadj:2019kji,Folacci:2019cmc,Folacci:2019vtt}. The Debye components
\eqref{eq:f_Debye_sum} (see also \eqref{eq:f_p_def_a} and \eqref{eq:f_p_def_b}) are obtained by adapting the same strategy.

For each real frequency $\omega$ and integer multipole $\ell$, we construct the exterior basis solutions by integrating the homogeneous Regge-Wheeler Eq.~\eqref{eq:radial} for $r\ge R$. The integration is initialized at large radius from the ingoing/outgoing asymptotic expansions, which are Pad\'e-resummed to improve accuracy at finite radius, and the solutions are then integrated down to the surface $r=R$. Imposing the matching conditions at $r=R$ with the local interior surface basis $\{u^{\rm in}_{\omega\ell},u^{\rm out}_{\omega\ell}\}$ defined in \eqref{eq:uin_uout_surface} yields the surface connection coefficients (in particular $\alpha^{\rm in}_\ell$, $\beta^{\rm out}_\ell$, and $\delta^{\rm in}_\ell$).

The internal phase factor $\xi_\ell(\omega)$ is determined from the interior problem. We integrate the regular interior solution from the center up to the surface $r=R$ (see also Sec.~\ref{subsec:numerics}) and match it there onto the local interior basis. This matching provides the coefficients $c^{\rm in}_\ell(\omega)$ and $c^{\rm out}_\ell(\omega)$, and hence $\xi_\ell(\omega)$. Once the surface connection coefficients and $\xi_\ell(\omega)$ are known, the $S$-matrix elements $S_\ell^{(0)}(\omega)$ and $S_\ell^{(p)}(\omega)$ are constructed and then used in Eqs.~\eqref{eq:f_p_def_a} and \eqref{eq:f_p_def_b} to compute the Debye scattering amplitudes.

A technical point concerns the square root in $k_{\rm int}$, Eq.~\eqref{eq:kint_def}, which introduces a threshold at $\ell=\ell_c$ defined by $k_{\rm int}(\omega,\ell_c)=0$ (i.e.\ $\omega^2=V_\ell(R^-)$).
Throughout this work we adopt the principal branch, chosen such that $\Im(k_{\rm int})>0$, which yields a smooth continuation between propagating ($k_{\rm int}\in\mathbb{R}$) and evanescent ($k_{\rm int}=i\kappa$) interior behavior and provides a consistent definition of the local interior basis for all integer $\ell$. For $\ell>\ell_c$ the interior is evanescent at the surface and transmission through the surface is strongly suppressed; consequently, the Debye orders $p\ge1$ become rapidly negligible (often within numerical precision). This explains why, in practice, the partial-wave sums for $p\ge1$ converge with significantly smaller truncation indices than the full amplitude or the $p=0$ contribution.

Due to the long-range nature of wave propagation in the Schwarzschild exterior, the partial-wave series for the full scattering amplitude \eqref{eq:pw_series} and the $p=0$ Debye contribution \eqref{eq:f_p_def_a} both suffer from poor convergence unless a convergence acceleration procedure is employed. To remedy this issue we use the method described in the Appendix of Ref.~\cite{Folacci:2019cmc}. By contrast, this acceleration is not required for the higher Debye orders $p\ge1$ [Eq.~\eqref{eq:f_p_def_b}], whose partial-wave sums converge much more rapidly once the long-range part carried by the direct ($p=0$) contribution has been separated out.

Accordingly, the partial-wave sums are truncated at some $\ell_{\max}$ after convergence tests. Here the ``full'' result refers to the direct evaluation of the partial-wave series \eqref{eq:pw_series} without Debye decomposition. In practice, the full amplitude and the $p=0$ contribution require substantially larger values of $\ell_{\max}$ than the higher Debye orders $p\ge1$. For the neutron star-like object ($R=6M$), we use $\ell_{\max}=55$ for the full series and for $p=0$ at $2M\omega=2$, while $\ell_{\max}=8$ is sufficient for $p\ge1$; at $2M\omega=16$, the corresponding truncations are $\ell_{\max}=300$ (full and $p=0$) and $\ell_{\max}=60$ ($p\ge1$). For the ultracompact object ($R=2.26M$), we take $\ell_{\max}=55$ for the full series and for $p=0$ at $2M\omega=2$ and $\ell_{\max}=8$ for $p\ge1$, whereas at $2M\omega=6$ we use $\ell_{\max}=120$ (full and $p=0$) and $\ell_{\max}=20$ ($p\ge1$). All numerical calculations were performed using {\it Mathematica}~\cite{Mathematica}.

%---------------------------------------------------------
\subsection{Results: Debye-series reconstruction of the scattering cross section}
\label{subsec:debye_scatt_results}

\begin{figure*}[htb]
	\centering
	\includegraphics[scale=0.52]{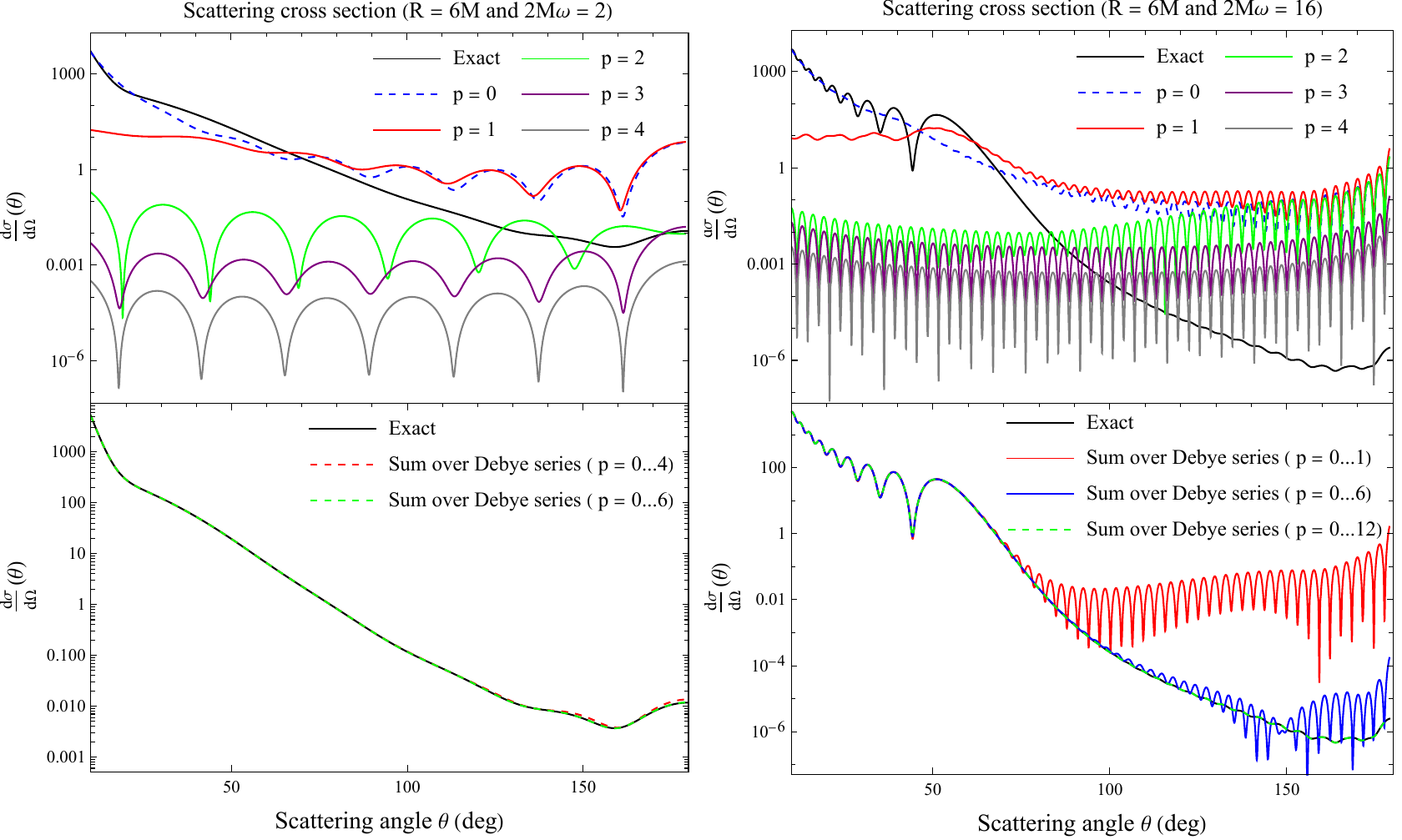}
	\caption{Differential scattering cross section for a neutron star-like compact body with $R=6M$.
			Upper panels: exact result (black curve) and the first Debye-order contributions $p=0,\ldots,4$ for $2M\omega=2$ (left) and $2M\omega=16$ (right).
			Lower panels: Debye partial sums compared with the exact cross section. At $2M\omega=2$ (left) only a few Debye orders are required, whereas at $2M\omega=16$ (right) higher orders become increasingly important, especially at large scattering angles.}
	\label{fig:Debye_vs_exact_R6}
\end{figure*}
\begin{figure*}[htb]
	\centering
	\includegraphics[scale=0.52]{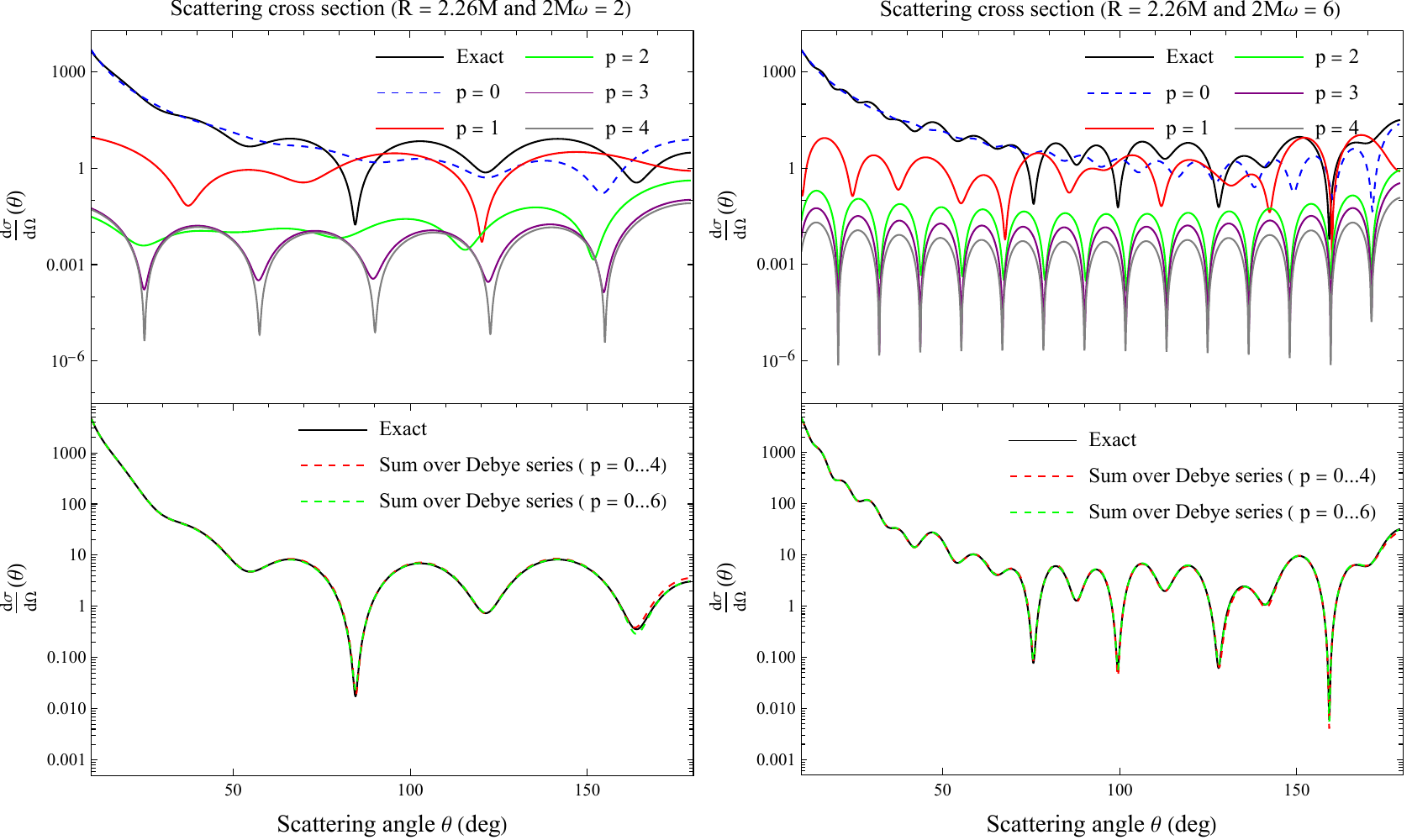}
	\caption{Differential scattering cross section for an ultracompact object with $R=2.26M$.
			Upper panels: exact result (black curve) and the first Debye-order contributions $p=0,\ldots,4$ for $2M\omega=2$ (left) and $2M\omega=6$ (right).
			Lower panels: Debye partial sums compared with the exact cross section. For this ultracompact configuration, a small number of Debye orders already yields an accurate approximation over a broad angular range at both frequencies.}
	\label{fig:Debye_vs_exact_R226}
\end{figure*}

Figures~\ref{fig:Debye_vs_exact_R6} and \ref{fig:Debye_vs_exact_R226} display representative results for the differential scattering cross section $\dd\sigma/\dd\Omega$ obtained from the Debye-series decomposition, and compare them with the standard partial-wave computation. In the upper panels we show the individual Debye-order contributions through $\bigl|f^{(p)}(\omega,\theta)\bigr|^2$, while the lower panels display reconstructed cross sections from partial Debye sums. We consider two configurations: a neutron star-like object with $R/M=6$ and an ultracompact object with $R/M=2.26$.

Figure~\ref{fig:Debye_vs_exact_R6} shows, in the upper panels, the first Debye-order contributions ($p=0,\ldots,4$) together with the corresponding result obtained from the full partial-wave computation.
One observes that small-angle scattering is dominated by the $p=0$ term (direct surface reflection), while intermediate angles receive their main contributions from the two lowest orders $p=0$ and $p=1$.
At larger scattering angles, in the backward/glory region, the cross section results from the interference of several Debye orders, producing the characteristic oscillatory pattern. The lower panels
illustrate the convergence of the reconstruction. In the intermediate-frequency regime $2M\omega=2$ (left), only a few Debye orders are needed: in practice, summing the first six terms already provides an
essentially complete reconstruction of the cross section for the neutron star-like model ($R/M=6$). At the higher frequency $2M\omega=16$ (right), the convergence becomes slower and additional Debye orders
are required (up to about $p\simeq 12$) to recover the fine oscillatory structure, especially at large scattering angles.

It is also noteworthy that, contrary to the classical Mie-scattering setting where rainbow features are typically associated with higher Debye orders (with the primary rainbow arising at $p=2$)~\cite{Nussenzveig:2006}, the present compact-body problem already exhibits a dominant  rainbow-like structure carried by the $p=1$ contribution (see the upper right panel for $2M\omega=16$). While the overall amplitude requires the inclusion of $p=0$, the $p=1$ term largely generates the sequence of peaks and troughs around the rainbow angle $\theta_r\simeq59.6^\circ$~\cite{Stratton:2019deq}. This indicates that a single interior passage through the effective potential may already produce the relevant caustic/interference structure, whereas higher orders $p\ge2$ mainly provide additional multibounce contributions and supernumerary interference.

Figure~\ref{fig:Debye_vs_exact_R226} shows the differential scattering cross section for the ultracompact configuration $R=2.26M$, together with the first Debye-order contributions ($p=0,\ldots,4$), for
$2M\omega=2$ (left) and $2M\omega=6$ (right). The upper panels indicate that, in both frequency regimes, the dominant contributions arise from the two lowest orders, $p=0$ and $p=1$, while higher orders are strongly suppressed. The lower panels show that summing only a small number of Debye orders already yields a very accurate reconstruction of the differential cross section over a broad angular range. The rapid convergence suggests that contributions associated with repeated interior propagation are more efficiently damped in the ultracompact case.

In this ultracompact regime, the deflection function can admit a rainbow extremum beyond the backward direction (i.e.\ at deflection angles exceeding $\pi$)~\cite{Stratton:2019deq}. The cross section displays fairly regular orbital oscillations as a function of $\theta$, with an angular scale set by the light-ring barrier.

%==============================================================
\section{CAM representation of Debye amplitudes}
\label{sec:CAM_debye}

In this section we reconstruct, by means of CAM techniques, the scattering amplitudes associated with each Debye order, and we compare these CAM reconstructions with the corresponding Debye-series results presented in Sec.~\ref{sec:results_debye}.

%------------------------------------------------------------
\subsection{Sommerfeld-Watson representation of the scattering amplitude}
\label{sec:CAM_debye_1}

The Sommerfeld-Watson (SW) transform converts a discrete partial-wave sum into a contour integral in the complex angular momentum plane (see~\cite{Newton:1982qc,Watson18,Sommerfeld49}). When the relevant function $F$ admits an analytic continuation that is regular on the real $\lambda$ axis, the standard SW procedure replaces the sum over integer $\ell$ by a contour integral in the complex variable $\lambda=\ell+1/2$.

\begin{figure}[htb]
	\centering
	\includegraphics[scale=0.50]{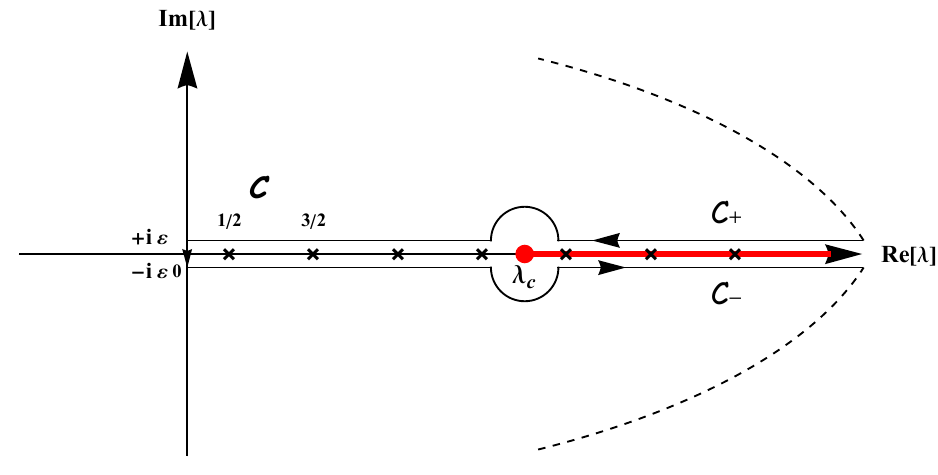}
	\caption{
		Sommerfeld-Watson contour $\mathcal{C}=\mathcal{C}_+\cup\mathcal{C}_-$ in the complex $\lambda$-plane in the presence of a branch cut starting at $\lambda_c$. The paths $\mathcal{C}_+$ and $\mathcal{C}_-$ run respectively just above and just below the real axis (prescriptions $\pm i\epsilon$), thereby probing the two limiting values $F^\pm$ on the lips of the cut. In the contour deformation, these paths are completed by the usual quarter circles at infinity in the first and fourth quadrants, together with a small keyhole contour around the branch point at $\lambda_c$. This deformation allows one to isolate, by Cauchy's theorem, the contributions of the Regge-Debye poles enclosed in the upper half-plane, as well as the cut and background integrals.
	}
	\label{fig:Complex_counter}
\end{figure}

In the present problem, the analytic structure is more delicate. For integer $\ell$, the interior wavenumber $k_{\rm int}(\omega,\ell)$ introduces a propagation/evanescence threshold at $\ell=\ell_c$ through $k_{\rm int}(\omega,\ell_c)=0$, and our construction uses the principal square-root branch such that $\Im(k_{\rm int})>0$ in the evanescent regime. After analytic continuation to complex $\lambda$, the same square root makes $k_{\rm int}(\omega,\lambda)$ multivalued and produces a genuine branch point at $\lambda_c=\ell_c+1/2$, together with a branch cut along the real axis for $\lambda\ge\lambda_c$ (see Fig.~\ref{fig:Complex_counter}). As a consequence, the usual SW contour deformation must be adapted to account for the cut~\cite{Newton:1982qc} and to ensure that the final representation reproduces the partial-wave sum on the physical sheet.

Deforming the SW contour onto the real axis yields, in addition to the pole contributions at $\lambda=\ell+1/2$, a term supported on the two lips of the cut and involving the boundary values $F^{\pm}$. The regulators $\pm i\epsilon$ thus label the two limiting values of the multivalued function. In our setting the physical continuation corresponds to the lower lip, $F^{-}$ (i.e.\ $\lambda\to\lambda-i0$), selected by the boundary conditions and equivalently by the prescription $\Im(k_{\rm int})>0$ for $\lambda>\lambda_c$. The additional principal-value term removes the extra (unphysical sheet) piece that would otherwise remain, leaving an identity written solely in terms of $F^{-}$ and therefore reproducing the original discrete sum.

We thus employ the modified Sommerfeld-Watson transform
\begin{equation}
	\label{eq:SWT_final}
	\begin{aligned}
		\sum_{\ell=0}^{\infty} &(-1)^\ell F(\ell)
		=\frac{i}{2}\Bigl\{ \,
		\int_{\mathcal{C}} d\lambda\,
		\frac{F(\lambda-1/2)}{\cos(\pi\lambda)}
		\\
		&+\int_{\lambda_c}^{\infty} d\lambda
		\left[
		\frac{F^{+}(\lambda-1/2)}{\cos\bigl(\pi(\lambda+i\epsilon)\bigr)}
		+\frac{F^{-}(\lambda-1/2)}{\cos\bigl(\pi(\lambda-i\epsilon)\bigr)}
		\right]
		\\
		&-2\,\mathrm{PV}\int_{\lambda_c}^{\infty} d\lambda\,
		\frac{F^{-}(\lambda-1/2)}{\cos(\pi\lambda)}
		\Bigr\},
	\end{aligned}
\end{equation}
valid when $F$ has a branch cut along the real $\lambda$ axis for $\lambda\ge\lambda_c$, with
\begin{equation}
	F^{\pm}(\lambda-1/2)=\lim_{\epsilon\to0^{+}}F(\lambda-1/2\pm i\epsilon).
\end{equation}
In Eq.~\eqref{eq:SWT_final}, the first term is the usual SW contour integral over $\mathcal{C}$ encircling the poles at $\lambda=\ell+1/2$. The second term accounts for the branch-cut contribution through the boundary values $F^{\pm}$ on the lips of the cut. The last term, written as a Cauchy principal value, enforces the selection of the physical branch $F^{-}$ and ensures equivalence with the partial-wave sum.

Applying \eqref{eq:SWT_final} to the Debye partial-wave expansions \eqref{eq:f_Debye_sum}, i.e.\ to Eqs.~\eqref{eq:f_p_def_a} and \eqref{eq:f_p_def_b}, replaces the discrete sum over $\ell$ by contour integrals in the complex $\lambda$-plane. Using
\begin{equation}
	P_\ell(\cos\theta)=(-1)^\ell P_\ell(-\cos\theta),
\end{equation}
we obtain, for the $p=0$ contribution,
\begin{widetext}
	\begin{equation}
		\label{eq:SWT_f0}
		\begin{aligned}
			f^{(0)}(\omega,\theta)
			=&{}\frac{1}{2\omega}\int_{\mathcal C} d\lambda\,
			\frac{\lambda}{\cos(\pi\lambda)}
			\left[S_{\lambda-1/2}^{(0)}(\omega)-1\right]
			P_{\lambda-1/2}(-\cos\theta) \\
			&\hspace{50pt} +\frac{1}{2\omega}\int_{\lambda_c}^{\infty} d\lambda\,\lambda
			\Biggl[
			\frac{S_{\lambda-1/2}^{(0),+}(\omega)-1}{\cos\bigl(\pi(\lambda+i\epsilon)\bigr)}
			+\frac{S_{\lambda-1/2}^{(0),-}(\omega)-1}{\cos\bigl(\pi(\lambda-i\epsilon)\bigr)}
			\Biggr]
			P_{\lambda-1/2}(-\cos\theta)
			\\
			&\hspace{50pt} -\frac{1}{\omega}\,\mathrm{PV}\int_{\lambda_c}^{\infty} d\lambda\,
			\frac{\lambda}{\cos\bigl(\pi\lambda\bigr)}
			\left[S_{\lambda-1/2}^{(0),-}(\omega)-1\right]
			P_{\lambda-1/2}(-\cos\theta).
		\end{aligned}
	\end{equation}
	Similarly, for $p\ge1$,
	\begin{equation}
		\label{eq:SWT_fp}
		\begin{aligned}
			f^{(p)}(\omega,\theta)
			=&{}\frac{1}{2\omega}\int_{\mathcal C} d\lambda\,
			\frac{\lambda}{\cos(\pi\lambda)}
			S_{\lambda-1/2}^{(p)}(\omega)\,
			P_{\lambda-1/2}(-\cos\theta)
			\\
			&\hspace{50pt}+\frac{1}{2\omega}\int_{\lambda_c}^{\infty} d\lambda\,\lambda
			\left[
			\frac{S_{\lambda-1/2}^{(p),+}(\omega)}{\cos\bigl(\pi(\lambda+i\epsilon)\bigr)}
			+\frac{S_{\lambda-1/2}^{(p),-}(\omega)}{\cos\bigl(\pi(\lambda-i\epsilon)\bigr)}
			\right]
			P_{\lambda-1/2}(-\cos\theta)
			\\
			&\hspace{50pt}-\frac{1}{\omega}\,\mathrm{PV}\int_{\lambda_c}^{\infty} d\lambda\,
			\frac{\lambda}{\cos\bigl(\pi\lambda\bigr)}
			S_{\lambda-1/2}^{(p),-}(\omega)\,
			P_{\lambda-1/2}(-\cos\theta).
		\end{aligned}
	\end{equation}
\end{widetext}					

In Eqs.~\eqref{eq:SWT_final}, \eqref{eq:SWT_f0}, and \eqref{eq:SWT_fp}, $\mathcal{C}$ denotes the keyhole-type Sommerfeld-Watson contour shown in Fig.~\ref{fig:Complex_counter}. It runs just above and below the real $\lambda$ axis (prescriptions $\pm i\epsilon$), is closed by a short segment joining $+i\epsilon$ to $-i\epsilon$, and is indented around the branch point at $\lambda=\lambda_c$.

The Legendre function of the first kind $P_{\lambda-1/2}(z)$ denotes the analytic continuation of the Legendre polynomials $P_\ell(z)$ to complex angular momentum. It is defined in terms of the hypergeometric function as~\cite{AS65}
\begin{equation}
	\label{eq:Legendre_Hypergeometric}
	P_{\lambda-1/2}(z)
	={}_2F_1\!\left(\frac12-\lambda,\frac12+\lambda;1;\frac{1-z}{2}\right).
\end{equation}

In Eqs.~\eqref{eq:SWT_f0} and \eqref{eq:SWT_fp}, $S_{\lambda-1/2}^{(0)}(\omega)$ and $S_{\lambda-1/2}^{(p)}(\omega)$ denote the analytic continuations of $S_{\ell}^{(0)}(\omega)$ and $S_{\ell}^{(p)}(\omega)$, respectively. Using Eqs.~\eqref{eq:S_Debye_terms_coeffs_a} and \eqref{eq:S_Debye_terms_coeffs_b}, they are given by
\begin{subequations}\label{eq:S_Debye_terms_coeffs_complex}
	\begin{align}
		S^{(0)}_{\lambda-1/2}(\omega)
		&=e^{i(\lambda+1/2)\pi}\,
		\frac{\beta^{\rm out}_{\lambda-1/2}(\omega)}{\alpha^{\rm in}_{\lambda-1/2}(\omega)},
		\label{eq:S_Debye_terms_coeffs_a_complex}
		\\
		S^{(p)}_{\lambda-1/2}(\omega)
		&=e^{i(\lambda+1/2)\pi}\,
		\left(\frac{k_{\rm int}(\omega,\lambda-1/2)}{\omega}\right)^{p}\nonumber\\
		&\times
		\frac{\xi_{\lambda-1/2}^{\,p}(\omega)\,\Bigl(\delta^{\rm in}_{\lambda-1/2}(\omega)\Bigr)^{p-1}}
		{\Bigl(\alpha^{\rm in}_{\lambda-1/2}(\omega)\Bigr)^{p+1}},
		\qquad p\ge1.
		\label{eq:S_Debye_terms_coeffs_b_complex}
	\end{align}
\end{subequations}
Here the complex surface connection coefficients $\alpha^{\rm in}_{\lambda-1/2}(\omega)$, $\beta^{\rm out}_{\lambda-1/2}(\omega)$, and $\delta^{\rm in}_{\lambda-1/2}(\omega)$, as well as the interior phase factor $\xi_{\lambda-1/2}(\omega)$, are defined from the analytic continuation of the radial basis functions and of the regular interior solution, namely from $u^{\rm in/out}_{\omega,\lambda-1/2}$, $f^{\rm in/up}_{\omega,\lambda-1/2}$, and $\phi^{\rm reg}_{\omega,\lambda-1/2}$. The interior wavenumber is continued as
\begin{equation}
	k_{\rm int}(\omega,\lambda-1/2)=\sqrt{\omega^2-V_{\lambda-1/2}(R^-)}.
\end{equation}
For $\lambda\ge\lambda_c$, we denote by $S^{(0),\pm}_{\lambda-1/2}(\omega)$ and $S^{(p),\pm}_{\lambda-1/2}(\omega)$ the limiting values of these analytically continued Debye $S$-matrix elements on the upper and lower lips of the branch cut, i.e.\ for $\lambda\to\lambda\pm i0$.

It is useful to recall the pole structure of the analytically continued Debye $S$-matrix elements in the complex $\lambda$-plane. The poles of $S^{(0)}_{\lambda-1/2}(\omega)$ (the Regge-Debye poles associated with the $p=0$ contribution) lie in the first and third quadrants and are symmetrically distributed with respect to the origin. From Eq.~\eqref{eq:S_Debye_terms_coeffs_a_complex}, they correspond to the zeros of the surface coefficient $\alpha^{\rm in}_{\lambda-1/2}(\omega)$, i.e.\ to the solutions $\lambda=\lambda_n(\omega)$ of
\begin{equation}
	\alpha^{\rm in}_{\lambda_n(\omega)-1/2}(\omega)=0,
	\qquad n=1,2,3,\dots .
\end{equation}

For $p\ge1$, the pole structure of $S^{(p)}_{\lambda-1/2}(\omega)$ is richer. Besides the poles associated with the zeros of $\alpha^{\rm in}_{\lambda-1/2}(\omega)$ [see Eq.~\eqref{eq:S_Debye_terms_coeffs_b_complex}], additional poles may occur through the internal phase factor $\xi_{\lambda-1/2}(\omega)$, i.e.\ at the zeros of $c^{\rm in}_{\lambda-1/2}(\omega)$ [see Eq.~\eqref{eq:xi_def}],
\begin{equation}
	c^{\rm in}_{\lambda_n(\omega)-1/2}(\omega)=0.
	\qquad n=1,2,3,\dots .
\end{equation}

Before deforming the contour, it is convenient to introduce the residues associated with the poles of the analytically continued Debye $S$-matrix elements, which play a central role in the CAM paradigm. Since the CAM expressions \eqref{eq:SWT_f0} and \eqref{eq:SWT_fp} involve the full integrands, we also define the corresponding CAM residues, which include the factor
\begin{equation}
	\mathcal{Q}(\lambda,\theta)\equiv \frac{\lambda}{\cos(\pi\lambda)}\,P_{\lambda-1/2}(-\cos\theta).
\end{equation}

For the $p=0$ contribution, the poles of $S^{(0)}_{\lambda-1/2}(\omega)$ arise from the zeros of $\alpha^{\rm in}_{\lambda-1/2}(\omega)$ [see Eq.~\eqref{eq:S_Debye_terms_coeffs_a_complex}]. Assuming a simple zero at $\lambda=\lambda_n^{(\alpha)}(\omega)$, the residue of the $S$-matrix element is
\begin{equation}
	r_n^{(0)}(\omega)
	=
	e^{i\pi[\lambda_n^{(\alpha)}(\omega)+1/2]}\,
	\frac{\beta^{\rm out}_{\lambda-1/2}(\omega)}
	{\displaystyle \frac{d}{d\lambda}\alpha^{\rm in}_{\lambda-1/2}(\omega)}
	\Big|_{\lambda=\lambda_n^{(\alpha)}(\omega)} .
\end{equation}

For $p\ge1$, $S^{(p)}_{\lambda-1/2}(\omega)$ receives pole contributions from the zeros of $\alpha^{\rm in}_{\lambda-1/2}(\omega)$ and from the zeros of $c^{\rm in}_{\lambda-1/2}(\omega)$. We treat these two cases separately.

(i) Poles at $\alpha^{\rm in}_{\lambda-1/2}(\omega)=0$. From Eq.~\eqref{eq:S_Debye_terms_coeffs_b_complex}, if $\lambda=\lambda_n^{(\alpha)}(\omega)$ is a simple zero of $\alpha^{\rm in}_{\lambda-1/2}(\omega)$ which is not simultaneously a zero of $c^{\rm in}_{\lambda-1/2}(\omega)$, then $S^{(p)}_{\lambda-1/2}(\omega)$ has a pole of order $p+1$. It is convenient to write
\begin{equation}
	S^{(p)}_{\lambda-1/2}(\omega)
	=
	\frac{G_p^{(\alpha)}(\lambda,\omega)}
	{\Bigl(\alpha^{\rm in}_{\lambda-1/2}(\omega)\Bigr)^{p+1}},
\end{equation}
with
\begin{align}
	G_p^{(\alpha)}(\lambda,\omega)
	={}&e^{i(\lambda+1/2)\pi}
	\left(\frac{k_{\rm int}(\omega,\lambda-1/2)}{\omega}\right)^p
	\nonumber\\
	&\times
	\Bigl(\delta^{\rm in}_{\lambda-1/2}(\omega)\Bigr)^{p-1}
	\Bigl(\xi_{\lambda-1/2}(\omega)\Bigr)^p .
\end{align}
We then define
\begin{align}
	h_n^{(\alpha)}(\lambda,\omega)
	&\equiv
	\frac{\alpha^{\rm in}_{\lambda-1/2}(\omega)}
	{\lambda-\lambda_n^{(\alpha)}(\omega)},
	\\
	h_n^{(\alpha)}\bigl(\lambda_n^{(\alpha)}(\omega),\omega\bigr)
	&=
	\left.\frac{d}{d\lambda}\alpha^{\rm in}_{\lambda-1/2}(\omega)\right|_{\lambda=\lambda_n^{(\alpha)}(\omega)} .
\end{align}
More generally, from the Taylor expansion of $\alpha^{\rm in}_{\lambda-1/2}(\omega)$ about $\lambda=\lambda_n^{(\alpha)}(\omega)$,
\begin{equation}
	\small
	\left.
	\frac{d^k}{d\lambda^k}h_n^{(\alpha)}(\lambda,\omega)
	\right|_{\lambda=\lambda_n^{(\alpha)}(\omega)}
	=
	\frac{1}{k+1}
	\left.
	\frac{d^{k+1}}{d\lambda^{k+1}}\alpha^{\rm in}_{\lambda-1/2}(\omega)
	\right|_{\lambda=\lambda_n^{(\alpha)}(\omega)}\!\!\!\!\!,
\end{equation}
with $k=0,1,2,\dots$.

The residue of $S^{(p)}_{\lambda-1/2}(\omega)$ at $\lambda=\lambda_n^{(\alpha)}(\omega)$ is
\begin{equation}
	r_{n,\alpha}^{(p)}(\omega)
	=
	\frac{1}{p!}
	\left[
	\frac{d^{\,p}}{d\lambda^{p}}
	\left(
	\frac{G_p^{(\alpha)}(\lambda,\omega)}
	{\bigl(h_n^{(\alpha)}(\lambda,\omega)\bigr)^{p+1}}
	\right)
	\right]_{\lambda=\lambda_n^{(\alpha)}(\omega)} .
\end{equation}
The corresponding CAM residue entering the Regge-Debye-pole contribution to $f^{(p)}(\omega,\theta)$ is
\begin{equation}
	\small
	\mathfrak{r}_{n,\alpha}^{(p)}(\omega,\theta)
	=
	\frac{1}{p!}
	\left[
	\frac{d^{\,p}}{d\lambda^{p}}
	\left(
	\frac{\mathcal{Q}(\lambda,\theta)\,G_p^{(\alpha)}(\lambda,\omega)}
	{\bigl(h_n^{(\alpha)}(\lambda,\omega)\bigr)^{p+1}}
	\right)
	\right]_{\lambda=\lambda_n^{(\alpha)}(\omega)} \!\!\!\!\!.
\end{equation}

(ii) Poles at $c^{\rm in}_{\lambda-1/2}(\omega)=0$. Using Eq.~\eqref{eq:xi_def}, Eq.~\eqref{eq:S_Debye_terms_coeffs_b_complex} can be rewritten as
\begin{equation}
	S^{(p)}_{\lambda-1/2}(\omega)
	=
	\frac{G_p^{(c)}(\lambda,\omega)}
	{\Bigl(c^{\rm in}_{\lambda-1/2}(\omega)\Bigr)^p},
\end{equation}
with
\begin{align}
	G_p^{(c)}(\lambda,\omega)
	={}&e^{i(\lambda+1/2)\pi}
	\left(\frac{k_{\rm int}(\omega,\lambda-1/2)}{\omega}\right)^p
	\nonumber\\
	&\times
	\frac{\Bigl(c^{\rm out}_{\lambda-1/2}(\omega)\Bigr)^p
		\Bigl(\delta^{\rm in}_{\lambda-1/2}(\omega)\Bigr)^{p-1}}
	{\Bigl(\alpha^{\rm in}_{\lambda-1/2}(\omega)\Bigr)^{p+1}} .
\end{align}
If $\lambda=\lambda_n^{(c)}(\omega)$ is a simple zero of $c^{\rm in}_{\lambda-1/2}(\omega)$ which is not simultaneously a zero of $\alpha^{\rm in}_{\lambda-1/2}(\omega)$, then $S^{(p)}_{\lambda-1/2}(\omega)$ has a pole of order $p$. We define
\begin{align}
	h_n^{(c)}(\lambda,\omega)
	&\equiv
	\frac{c^{\rm in}_{\lambda-1/2}(\omega)}
	{\lambda-\lambda_n^{(c)}(\omega)},
	\\
	h_n^{(c)}\bigl(\lambda_n^{(c)}(\omega),\omega\bigr)
	&=
	\left.\frac{d}{d\lambda}c^{\rm in}_{\lambda-1/2}(\omega)\right|_{\lambda=\lambda_n^{(c)}(\omega)} .
\end{align}
More generally,
\begin{equation}
	\small
	\left.
	\frac{d^k}{d\lambda^k}h_n^{(c)}(\lambda,\omega)
	\right|_{\lambda=\lambda_n^{(c)}(\omega)}
	=
	\frac{1}{k+1}
	\left.
	\frac{d^{k+1}}{d\lambda^{k+1}}c^{\rm in}_{\lambda-1/2}(\omega)
	\right|_{\lambda=\lambda_n^{(c)}(\omega)} \!\!\!\!\!,
\end{equation}
with $k=0,1,2,\dots$.

The residue of $S^{(p)}_{\lambda-1/2}(\omega)$ at $\lambda=\lambda_n^{(c)}(\omega)$ is
\begin{equation}
	r_{n,c}^{(p)}(\omega)
	=
	\frac{1}{(p-1)!}
	\left[
	\frac{d^{\,p-1}}{d\lambda^{p-1}}
	\left(
	\frac{G_p^{(c)}(\lambda,\omega)}
	{\bigl(h_n^{(c)}(\lambda,\omega)\bigr)^p}
	\right)
	\right]_{\lambda=\lambda_n^{(c)}(\omega)} \!\!\!\!\!.
\end{equation}
The corresponding CAM residue entering the Regge-Debye pole contribution to $f^{(p)}(\omega,\theta)$ is
\begin{equation}
	\small
	\mathfrak{r}_{n,c}^{(p)}(\omega,\theta)
	=
	\frac{1}{(p-1)!}
	\left[
	\frac{d^{\,p-1}}{d\lambda^{p-1}}
	\left(
	\frac{\mathcal{Q}(\lambda,\theta)\,G_p^{(c)}(\lambda,\omega)}
	{\bigl(h_n^{(c)}(\lambda,\omega)\bigr)^p}
	\right)
	\right]_{\lambda=\lambda_n^{(c)}(\omega)} \!\!\!\!\!.
\end{equation}

%--------------------------------------------------------------
\subsection{CAM representation of the Debye amplitudes}
\label{subsec:cam_debye_rep}

\subsubsection{CAM representation of $p=0$}
\label{subsubsec:cam_p0}

We now deform the contour $\mathcal{C}$ in Eq.~\eqref{eq:SWT_f0} and use Cauchy's theorem to extract the contributions from the Regge-Debye poles associated with the zeros of $\alpha^{\rm in}_{\lambda-1/2}(\omega)$ in the first quadrant of the CAM plane. The derivation follows the standard contour deformation strategy of CAM theory, here applied to the modified Sommerfeld-Watson representation \eqref{eq:SWT_f0} in the presence of a branch cut and principal-value terms; see Ref.~\cite{Folacci:2019cmc} (in particular, Sec.~IIB.3). In practice, the contour $\mathcal{C}$ is decomposed as $\mathcal{C}=\mathcal{C}_+\cup\mathcal{C}_-$ and deformed into the first and fourth quadrants of the complex $\lambda$-plane. The deformation of $\mathcal{C}_+$ into the first quadrant allows one to collect the contributions of the Regge-Debye poles lying in this quadrant, together with a background integral along the imaginary axis from $+i\infty$ to $0$. Similarly, the deformation of $\mathcal{C}_-$ into the fourth quadrant yields the corresponding background contribution along the imaginary axis from $-i\infty$ to $0$ (see Fig.~\eqref{fig:Complex_counter}).  In addition, quarter-circle arcs at infinity are introduced in both quadrants. As in Ref.~\cite{Folacci:2019cmc}, the contributions from these arcs vanish and are therefore discarded.

As a result, the Debye amplitude of order $p=0$ can be written as the sum of three distinct CAM contributions,
\begin{equation}
	\label{eq:f0_CAM_decomposition_p0}
	f^{(0)}(\omega,\theta)
	=
	f^{(0)}_{\rm B}(\omega,\theta)
	+
	f^{(0)}_{\rm RDP}(\omega,\theta)
	+
	f^{(0)}_{\rm cut}(\omega,\theta),
\end{equation}
where the three terms correspond, respectively, to a background integral contribution, a discrete sum over Regge-Debye poles, and a contribution associated with the branch-cut structure of the scattering matrix in the complex angular momentum plane.

\begin{subequations}\label{CAM_Scalar_Scattering_amp_decomp_p0}
	\begin{equation}\label{CAM_Scalar_Scattering_amp_decomp_Background_p0}
		f^{(0)}_{\rm B}(\omega,\theta)
		=
		f^{(0)}_{\rm B,Re}(\omega,\theta)
		+
		f^{(0)}_{\rm B,Im}(\omega,\theta),
	\end{equation}
	with
	\begin{equation}\label{CAM_Scalar_Scattering_amp_decomp_Background_a_p0}
		f^{(0)}_{\rm B,Re}(\omega,\theta)
		=
		\frac{1}{\pi \omega}
		\int_{{\cal C}_{-}} d\lambda \,
		\lambda\, S_{\lambda -1/2}^{(0),-}(\omega)\,
		Q_{\lambda -1/2}(\cos \theta + i0),
	\end{equation}
	and
	\begin{eqnarray}\label{CAM_Scalar_Scattering_amp_decomp_Background_b_p0}
		&& f^{(0)}_{\rm B, Im}(\omega,\theta) = \frac{1}{2 \omega}\left(\int_{+i\infty}^{0} d\lambda \, \left[S_{\lambda -1/2}^{(0),+}(\omega) P_{\lambda-1/2} (-\cos \theta) \right. \right.\nonumber \\
		&&- \left.  \left. S_{-\lambda -1/2}^{(0),-}(\omega) e^{i \pi \left(\lambda+1/2\right)}P_{\lambda-1/2} (\cos \theta) \right] \frac{\lambda}{\cos (\pi \lambda) } \right).
	\end{eqnarray}
\end{subequations}
is a background integral contribution. The Regge-Debye pole contribution is
\begin{equation}\label{CAM_Scalar_Scattering_amp_decomp_RP_p0}
	f^{(0)}_{\rm RDP}(\omega,\theta)
	=
	-\frac{i \pi}{\omega}
	\sum_{n=1}^{+\infty}
	\frac{\lambda_n^{(\alpha)}(\omega)\, r_n^{(0)}(\omega)}
	{\cos[\pi \lambda_n^{(\alpha)}(\omega)]}
	P_{\lambda_n^{(\alpha)}(\omega) -1/2} (-\cos \theta),
\end{equation}
where the sum runs over the Regge-Debye poles in the first quadrant of the CAM plane. Finally, the branch-cut contribution (including the principal-value corrective term) is
\begin{align}\label{CAM_Scalar_Scattering_amp_decomp_Branch_cut_a_p0}
	f^{(0)}_{\rm cut}&(\omega,\theta)= 
	\frac{1}{2\omega}\int_{\lambda_c}^{\infty} d\lambda\,\lambda
	\Bigl[
	\frac{S_{\lambda-1/2}^{(0),+}(\omega)-1}{\cos\bigl(\pi(\lambda+i\epsilon)\bigr)} \nonumber \\
	&+\frac{S_{\lambda-1/2}^{(0),-}(\omega)-1}{\cos\bigl(\pi(\lambda-i\epsilon)\bigr)}
	\Bigr]
	P_{\lambda-1/2}(-\cos\theta)
	\nonumber \\
	&-\frac{1}{\omega}\,\mathrm{PV}\int_{\lambda_c}^{\infty} d\lambda\,
	\frac{\lambda}{\cos\bigl(\pi\lambda\bigr)} \nonumber \\
	&\qquad\qquad\times \left[S_{\lambda-1/2}^{(0),-}(\omega) -1\right]\,
	P_{\lambda-1/2}(-\cos\theta).
\end{align}

Equations~\eqref{eq:f0_CAM_decomposition_p0}--\eqref{CAM_Scalar_Scattering_amp_decomp_Branch_cut_a_p0} provide an exact CAM representation of the $p=0$ Debye amplitude, fully equivalent to the partial-wave expansion~\eqref{eq:f_p_def_a}.

\subsubsection{CAM representation of $p\geq 1$}

We now deform the contour $\mathcal C$ in Eq.~\eqref{eq:SWT_fp} and use Cauchy's theorem to extract the Regge-Debye pole terms associated with the analytically continued Debye element $S_{\lambda-1/2}^{(p)}(\omega)$ for $p\ge1$ [see Eq.~\eqref{eq:S_Debye_terms_coeffs_b_complex}]. The derivation follows the same CAM contour deformation strategy as for $p=0$, adapted to the modified Sommerfeld-Watson representation \eqref{eq:SWT_fp} (see Ref.~\cite{Folacci:2019cmc}). In contrast with the $p=0$ sector, the pole sum now involves two families: poles associated with the zeros of $\alpha^{\rm in}_{\lambda-1/2}(\omega)$ and additional poles generated by the zeros of $c^{\rm in}_{\lambda-1/2}(\omega)$. The corresponding CAM residues are denoted $\mathfrak r_{n,\alpha}^{(p)}(\omega,\theta)$ and $\mathfrak r_{n,c}^{(p)}(\omega,\theta)$.

As a result, the Debye amplitude of order $p\ge1$ can be written as
\begin{equation}
	\label{eq:fp_CAM_decomposition_p}
	f^{(p)}(\omega,\theta)
	=
	f^{(p)}_{\rm B}(\omega,\theta)
	+
	f^{(p)}_{\rm RDP}(\omega,\theta)
	+
	f^{(p)}_{\rm cut}(\omega,\theta).
\end{equation}

In Eq.~\eqref{eq:fp_CAM_decomposition_p}, the background contribution is decomposed as     
\begin{subequations}\label{eq:fp_background_decomposition_p}
	\begin{equation}
		\label{eq:fp_background_split_p}
		f^{(p)}_{\rm B}(\omega,\theta)
		=
		f^{(p)}_{\rm B,Re}(\omega,\theta)
		+
		f^{(p)}_{\rm B,Im}(\omega,\theta),
	\end{equation}
	with
	\begin{equation}
		\label{eq:fp_background_Re_p}
		f^{(p)}_{\rm B,Re}(\omega,\theta)
		=
		\frac{1}{\pi\omega}
		\int_{\mathcal C_-} d\lambda\,
		\lambda\,S_{\lambda-1/2}^{(p),-}(\omega)\,
		Q_{\lambda-1/2}(\cos\theta+i0),
	\end{equation} 
	and
	\begin{eqnarray}\label{CAM_Scalar_Scattering_amp_decomp_Background_b_p}
	&& f^{(p)}_{\rm B, Im}(\omega,\theta) = \frac{1}{2 \omega}\int_{+i\infty}^{0} d\lambda \, \left[S_{\lambda -1/2}^{(p),+}(\omega) P_{\lambda-1/2} (-\cos \theta) \right.\nonumber \\
	&&-  \left. S_{-\lambda -1/2}^{(p),-}(\omega) e^{i \pi \left(\lambda+1/2\right)}P_{\lambda-1/2} (\cos \theta) \right] \frac{\lambda}{\cos (\pi \lambda) }.
	\end{eqnarray}
\end{subequations}

The Regge-Debye pole contribution is        
\begin{equation}    
	\label{eq:fp_RDP_contribution_p}
	f^{(p)}_{\rm RDP}(\omega,\theta)
	=
	-\frac{i\pi}{\omega}
	\left[
	\sum_{n=1}^{\infty}\mathfrak r_{n,\alpha}^{(p)}(\omega,\theta)
	-
	\sum_{n=1}^{N}\mathfrak r_{n,c}^{(p)}(\omega,\theta)
	\right],
\end{equation}
where $N$ denotes the (finite) number of poles associated with the zeros of $c^{\rm in}_{\lambda-1/2}(\omega)$ that are retained in the first quadrant of the CAM plane for the configuration under consideration; it depends on the physical configuration (in particular on the radius $R$) and on the frequency $\omega$.

Finally, the cut contribution reads    
\begin{align}\label{CAM_Scalar_Scattering_amp_decomp_Branch_cut_a_p}
	f^{(p)}_{\rm cut}&(\omega,\theta)= 
	\frac{1}{2\omega}\int_{\lambda_c}^{\infty} d\lambda\,\lambda
	\Bigl[
	\frac{S_{\lambda-1/2}^{(p),+}(\omega)}{\cos\bigl(\pi(\lambda+i\epsilon)\bigr)} \nonumber \\
	&+\frac{S_{\lambda-1/2}^{(p),-}(\omega)}{\cos\bigl(\pi(\lambda-i\epsilon)\bigr)}
	\Bigr]
	P_{\lambda-1/2}(-\cos\theta)
	\nonumber \\
	&-\frac{1}{\omega}\,\mathrm{PV}\int_{\lambda_c}^{\infty} d\lambda\,
	\frac{\lambda}{\cos\bigl(\pi\lambda\bigr)}
	S_{\lambda-1/2}^{(p),-}(\omega)\,
	P_{\lambda-1/2}(-\cos\theta).
\end{align}

Likewise, Eqs.~\eqref{eq:fp_CAM_decomposition_p}--\eqref{CAM_Scalar_Scattering_amp_decomp_Branch_cut_a_p} provide the exact CAM representation of the Debye contribution of order $p\ge1$, equivalent to the partial-wave expansion~\eqref{eq:f_p_def_b}.	

%----------------------------------------------------
\subsection{Numerical computation of the CAM contributions}
\label{sec:CAM_debye_3}

The numerical evaluation of the CAM ingredients entering the Debye amplitudes, namely the Regge-Debye pole sums, the background integrals, and the branch-cut terms, follows the computational strategies developed in Refs.~\cite{Folacci:2019cmc,Folacci:2019vtt}, adapted to the present Debye-CAM framework. In practice, we compute these quantities for both the $p=0$ sector [Eqs.~\eqref{CAM_Scalar_Scattering_amp_decomp_Background_a_p0}--\eqref{CAM_Scalar_Scattering_amp_decomp_Branch_cut_a_p0}]
and the $p\ge1$ sector [Eqs.~\eqref{eq:fp_background_Re_p}--\eqref{CAM_Scalar_Scattering_amp_decomp_Branch_cut_a_p}].

The branch-cut contribution requires the explicit construction of the analytically continued Debye $S$-matrix elements $S_{\lambda-1/2}^{(p),\pm}(\omega)$ on the two lips of the cut (i.e.\ the limits $\lambda\to\lambda\pm i0$ for $\lambda\ge\lambda_c$). The Regge-Debye pole contribution is obtained from the corresponding CAM residues; for $p\ge1$ this includes both the poles generated by the zeros of $\alpha^{\rm in}_{\lambda-1/2}(\omega)$ and those associated with the zeros of $c^{\rm in}_{\lambda-1/2}(\omega)$.

From a numerical perspective, the real-axis background integrals \eqref{CAM_Scalar_Scattering_amp_decomp_Background_a_p0} and \eqref{eq:fp_background_Re_p} exhibit slow convergence and are evaluated using the same convergence-acceleration/regularization procedure as in Ref.~\cite{Folacci:2019cmc}. By contrast, no such treatment is required for the imaginary-axis background integrals \eqref{CAM_Scalar_Scattering_amp_decomp_Background_b_p0} and \eqref{CAM_Scalar_Scattering_amp_decomp_Background_b_p}, whose integrands decay exponentially along the integration contour.

%----------------------------------------------------
\subsection{Results: CAM Debye-series reconstruction}
\label{sec:CAM_debye_4}

\begin{figure*}[!ht]%[htb]
	\centering
	\includegraphics[scale=0.58]{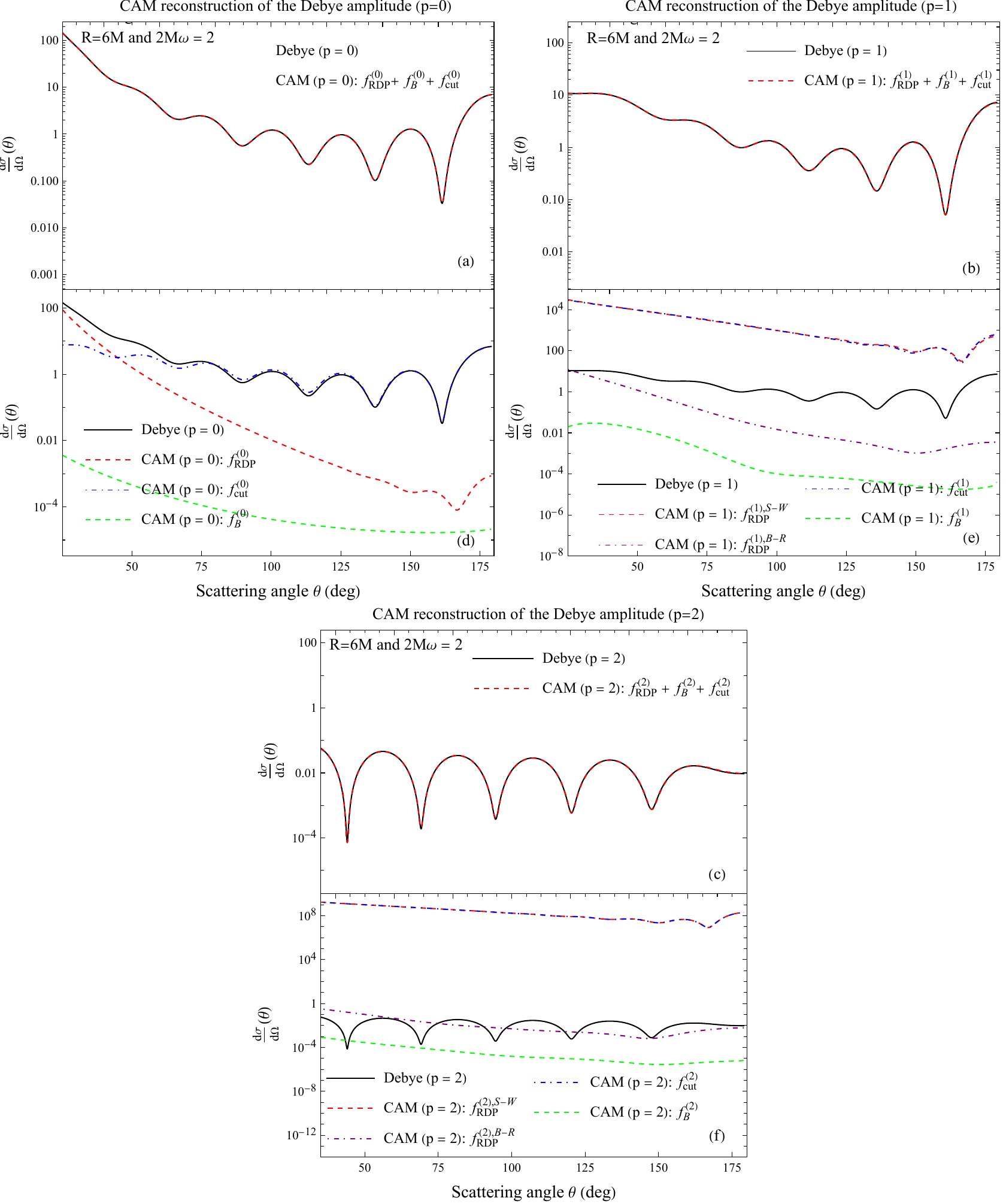}
	\caption{CAM reconstruction of Debye contributions for a neutron star-like compact body ($R=6M$, $2M\omega=2$).
		Upper panels: comparison between the Debye partial-wave result (black) and its CAM reconstruction (red dashed) for (a) $p=0$, (b) $p=1$, and (c) $p=2$. In these panels, the CAM reconstruction is given by $f^{(p)}_{\rm RDP}+f^{(p)}_{\rm B}+f^{(p)}_{\rm cut}$; although the background term is small on the plotted scale, it provides a noticeable correction at small scattering angles.
		Lower panels: CAM decomposition of the same Debye orders into Regge-Debye pole, branch-cut, and background contributions for (d) $p=0$, (e) $p=1$, and (f) $p=2$. For $p=1$ and $p=2$, the pole term is split into the surface-wave and broad-resonance families, corresponding to poles generated by $\alpha^{\rm in}$ and by $c^{\rm in}$, respectively. In the pole sums, we retain $N_{\rm S-W}=50$ surface-wave poles in all panels and $N_{\rm B-R}=3$ broad-resonance poles in the $p=1$ and $p=2$ reconstructions.}
	\label{fig:CAM_Debye_R6_2Mw2}
\end{figure*}
\begin{figure*}[!ht]%[htb]
	\centering
	\includegraphics[scale=0.58]{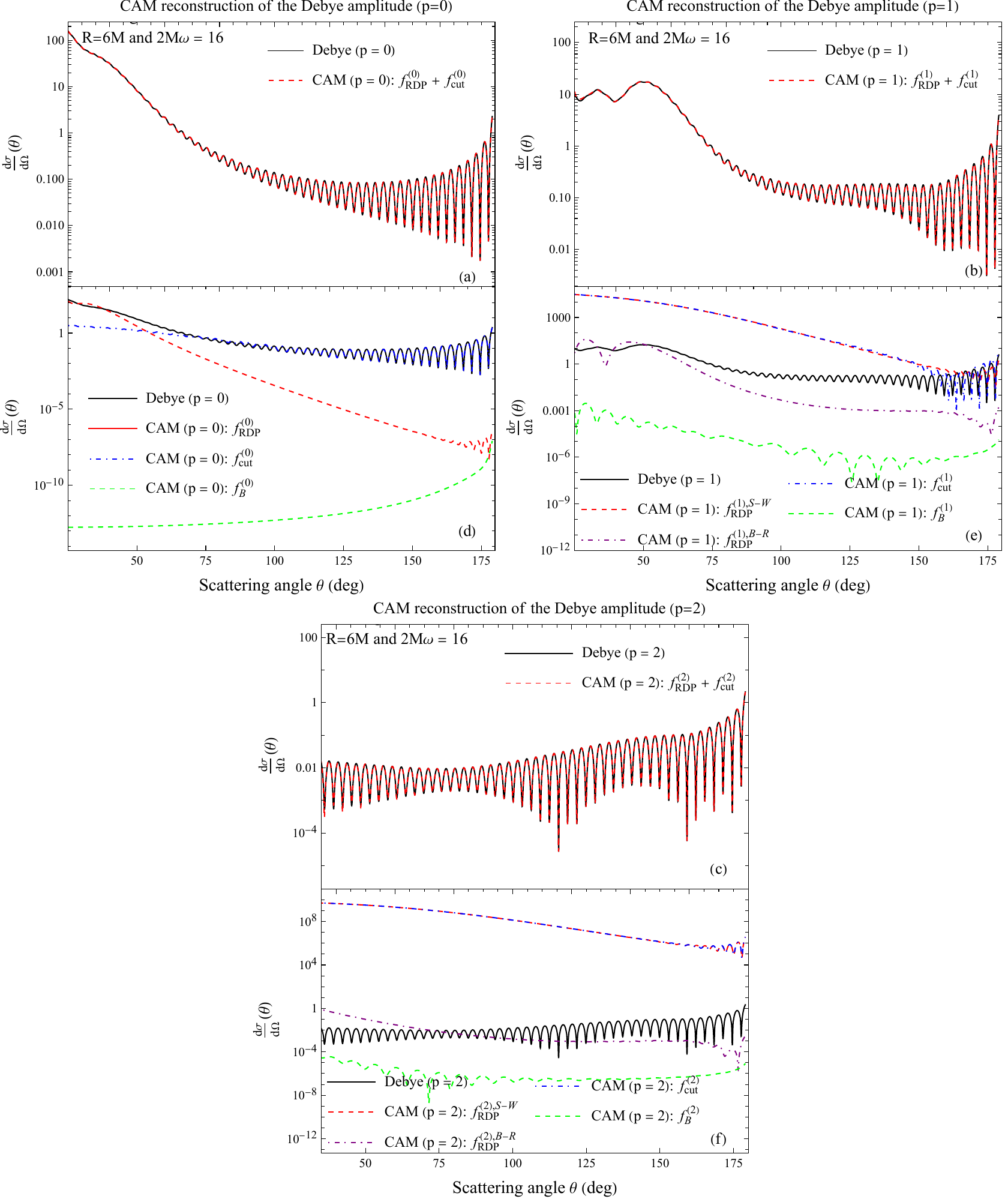}
	\caption{CAM reconstruction of Debye contributions for a neutron star-like compact body ($R=6M$, $2M\omega=16$).
		Upper panels: comparison between the Debye partial-wave result (black) and its CAM reconstruction (red dashed) for (a) $p=0$, (b) $p=1$, and (c) $p=2$. In these panels, the CAM reconstruction is displayed as $f^{(p)}_{\rm RDP}+f^{(p)}_{\rm cut}$ (the background term being numerically negligible on the plotted scale).
		Lower panels: CAM decomposition of the same Debye orders into Regge-Debye pole, branch-cut, and background contributions for (d) $p=0$, (e) $p=1$, and (f) $p=2$. For $p=1$ and $p=2$, the pole term is split into the surface-wave and broad-resonance families, corresponding to poles generated by $\alpha^{\rm in}$ and by $c^{\rm in}$, respectively. In the pole sums, we retain $N_{\rm S-W}=50$ surface-wave poles in all panels and $N_{\rm B-R}=22$ broad-resonance poles in the $p=1$ and $p=2$ reconstructions.}
	\label{fig:CAM_Debye_R6_2Mw16}
\end{figure*}
\begin{figure*}[!ht]%[htb]
	\centering
	\includegraphics[scale=0.58]{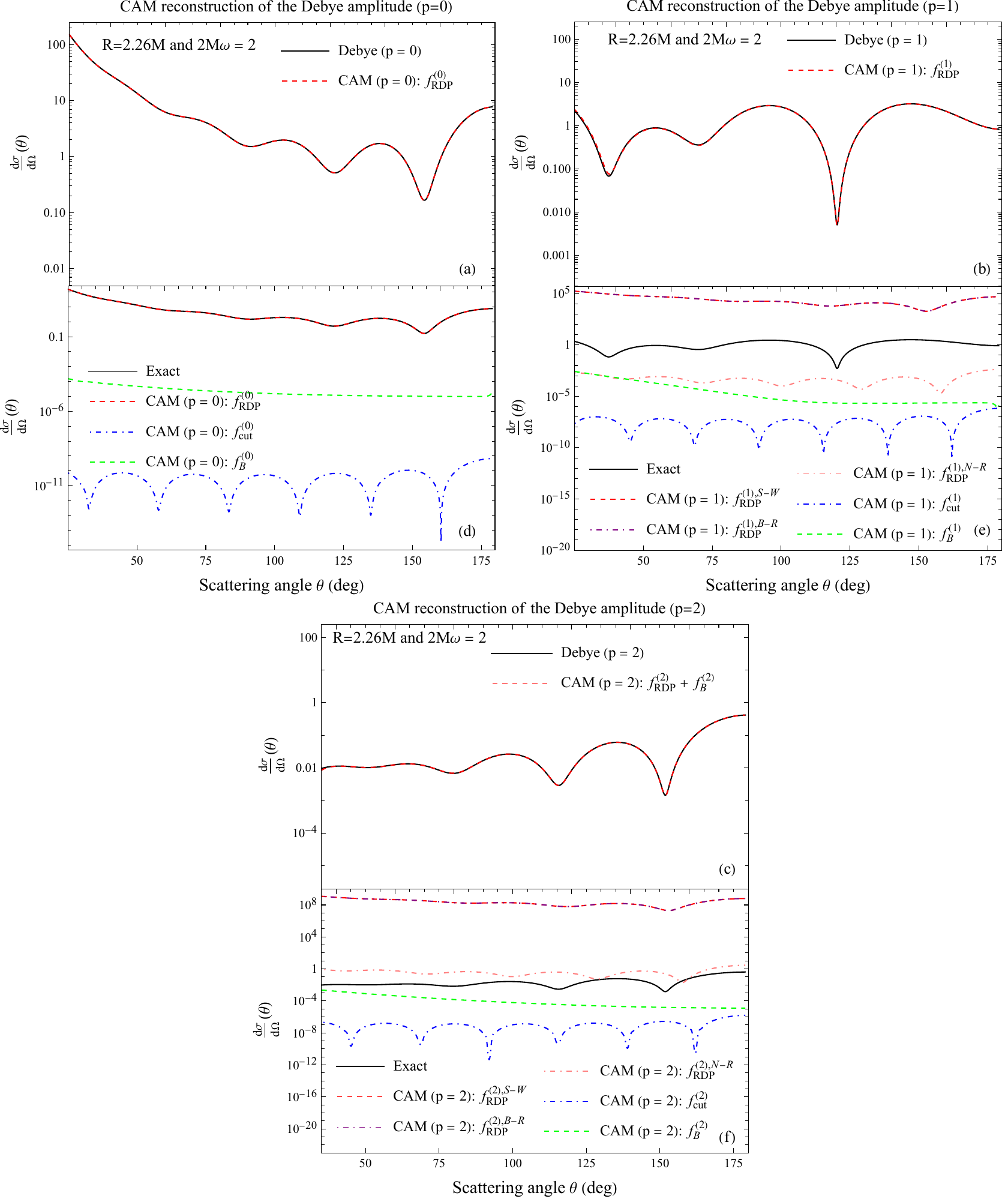}
	\caption{CAM reconstruction of the Debye contributions for an ultracompact object ($R=2.26M$, $2M\omega=2$).
		Upper panels: comparison between the Debye partial-wave result (black) and the CAM reconstruction (red dashed) for (a) $p=0$, (b) $p=1$, and (c) $p=2$. For $p=0$ and $p=1$ the reconstruction is obtained from the Regge-Debye pole sum only, i.e.\ $f^{(p)}_{\rm CAM}\simeq f^{(p)}_{\rm RDP}$, whereas for $p=2$ a small-angle correction from the background term is included, $f^{(2)}_{\rm CAM}=f^{(2)}_{\rm RDP}+f^{(2)}_{\rm B}$ (the cut term being negligible on the plotted scale).
		Lower panels: relative magnitude of the CAM components for the same Debye orders, (d) $p=0$, (e) $p=1$, and (f) $p=2$. For $p\ge1$, the pole contribution is split into the surface-wave family, generated by the zeros of $\alpha^{\rm in}$, and the interior-resonance poles generated by the zeros of $c^{\rm in}$, which separate into a broad-resonance and a narrow-resonance subfamily. In the pole sums, we retain $N_{\rm S-W}=50$ surface-wave poles in all panels, together with $N_{\rm B-R}=4$ broad-resonance poles and $N_{\rm N-R}=1$ narrow-resonance pole in the $p=1$ and $p=2$ reconstructions.}
	\label{fig:CAM_Debye_R226_2Mw2}
\end{figure*}
\begin{figure*}[!ht]%[htb]
	\centering
	\includegraphics[scale=0.58]{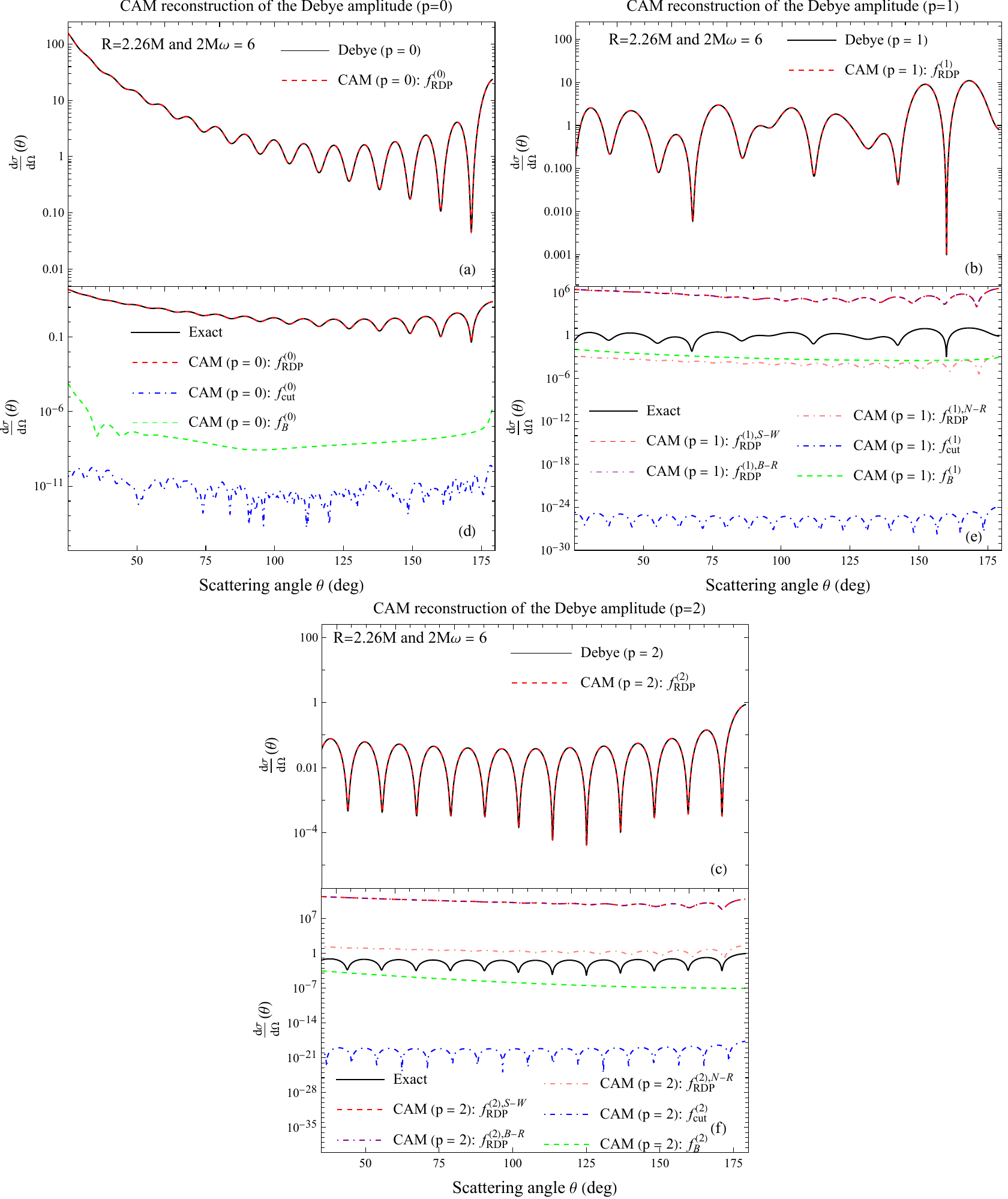}
	\caption{CAM reconstruction of the Debye contributions for an ultracompact object ($R=2.26M$, $2M\omega=6$).
		Upper panels: comparison between the Debye partial-wave result (black) and the CAM reconstruction obtained from the Regge-Debye pole sum only (red dashed) for (a) $p=0$, (b) $p=1$, and (c) $p=2$, i.e.\ $f^{(p)}_{\rm CAM}\simeq f^{(p)}_{\rm RDP}$.
		Lower panels: relative magnitude of the CAM components for the same Debye orders, (d) $p=0$, (e) $p=1$, and (f) $p=2$, showing that the background and branch-cut terms are negligible on the displayed scale. For $p\ge1$, the pole contribution is split into the surface-wave family, generated by the zeros of $\alpha^{\rm in}$, and the interior-resonance poles generated by the zeros of $c^{\rm in}$, which separate into a broad-resonance and a narrow-resonance subfamily. In the pole sums, we retain $N_{\rm S-W}=50$ surface-wave poles in all panels, together with $N_{\rm B-R}=11$ broad-resonance poles and $N_{\rm N-R}=3$ narrow-resonance pole in the $p=1$ and $p=2$ reconstructions.}
	\label{fig:CAM_Debye_R226_2Mw6}
\end{figure*}

In this subsection we present numerical results for the CAM reconstruction of the Debye amplitudes. For each Debye order $p$, the CAM representation expresses $f^{(p)}(\omega,\theta)$ as the sum of a Regge-Debye pole contribution, a background term, and a branch-cut term. We compare these CAM reconstructions with the corresponding Debye partial-wave results and use the CAM decomposition to quantify the relative importance of the pole, background, and cut sectors for neutron star-like ($R=6M$) and ultracompact ($R=2.26M$) configurations at representative frequencies.

In the numerical evaluation of the Regge-Debye pole contributions, the pole sums are truncated after convergence tests. In all configurations we retain $N_{\rm S-W}=50$ surface-wave poles (the family associated with the zeros of $\alpha^{\rm in}$), which is required to obtain an accurate reconstruction down to small scattering angles. The interior pole sector is handled differently: for the neutron star-like case we include all broad-resonance poles in the first quadrant for each frequency, while for the ultracompact case we also include the narrow-resonance poles when present. The corresponding numbers depend on the pair $(R,\omega)$.

For the neutron star-like configuration $R=6M$, Figs.~\ref{fig:CAM_Debye_R6_2Mw2} and \ref{fig:CAM_Debye_R6_2Mw16} validate the CAM reconstruction of the first Debye orders $p=0,1,2$ by direct comparison with the Debye partial-wave results (upper panels). In both frequency regimes the agreement is excellent for the dominant orders shown. The lower panels clarify the relative magnitude of the CAM ingredients. For $p=0$, the Regge-Debye pole term dominates at small and moderate angles, while the branch-cut term becomes comparable, and typically dominant at intermediate and large angles, i.e., toward the backward/glory region. The background term remains much smaller, although it can yield a visible correction at very small angles when included. For $p=1$ and $p=2$, the surface-wave pole contribution and the branch-cut term have comparable magnitudes. Their sum, $f^{(p),{\rm S\!-\!W}}_{\rm RDP}+f^{(p)}_{\rm cut}$, already reproduces $f^{(p)}$ in the backward/glory region. At intermediate and small angles, however, this two-term approximation is not sufficient: adding the broad-resonance pole sector $f^{(p),{\rm B\!-\!R}}_{\rm RDP}$ is necessary to recover the Debye result over the full angular range.

For the ultracompact configuration $R=2.26M$, Figs.~\ref{fig:CAM_Debye_R226_2Mw2} and \ref{fig:CAM_Debye_R226_2Mw6} show a markedly different situation: the Debye contributions are essentially pole dominated. In the upper panels, the pole sum alone already reproduces the Debye results extremely well for $p=0$ and $p=1$ at both frequencies. For $p=2$, a small background correction is required only at low frequency ($2M\omega=2$) and only at small scattering angles, whereas at $2M\omega=6$ the pole sum remains sufficient over the plotted range. The lower panels confirm that, in the ultracompact regime, both the branch-cut and background terms are suppressed by many orders of magnitude compared with the pole contribution. For $p\ge1$, the pole sector separates into the surface wave family associated with the zeros of $\alpha^{\rm in}$ and an interior-resonance sector associated with the zeros of $c^{\rm in}$. In the ultracompact case the latter splits into broad-resonance and narrow-resonance branches, with the narrow-resonance poles lying closer to the real axis.

\begin{figure}[!ht]%[htb]
	\centering
	\includegraphics[scale=0.60]{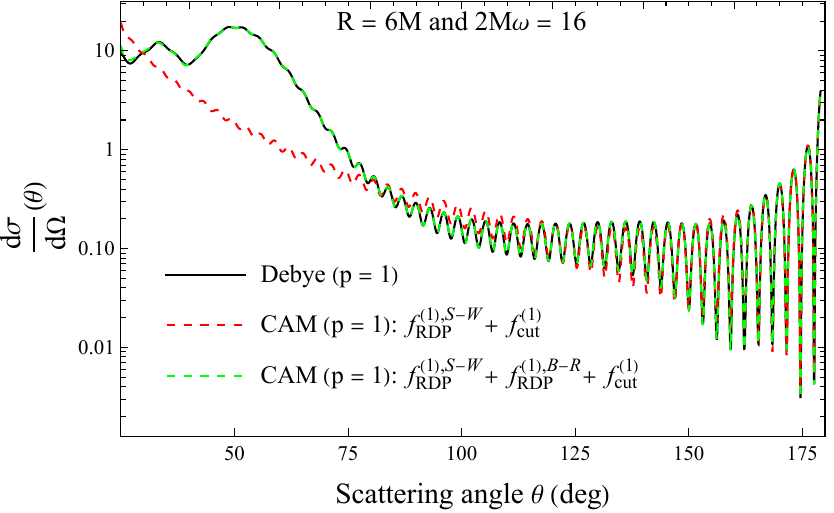}
	\caption{Role of the broad-resonance Regge-Debye poles in the $p=1$ CAM reconstruction for the neutron star-like configuration ($R=6M$, $2M\omega=16$).
		We compare the Debye partial-wave result for $p=1$ (black) with two partial CAM reconstructions: the sum of the surface-wave pole contribution and the branch-cut term, $f^{(1),{\rm S\!-\!W}}_{\rm RDP}+f^{(1)}_{\rm cut}$ (red dashed), and the same quantity supplemented by the broad-resonance poles, $f^{(1),{\rm S\!-\!W}}_{\rm RDP}+f^{(1),{\rm B\!-\!R}}_{\rm RDP}+f^{(1)}_{\rm cut}$ (green dashed). The broad-resonance sector is required to reproduce the rainbow enhancement around $\theta_r\simeq 59.6^\circ$, while $f^{(1),{\rm S\!-\!W}}_{\rm RDP}+f^{(1)}_{\rm cut}$ already captures the large-angle behavior.}
	\label{fig:CAM_Debye_R6_2Mw16_Rainbow}
\end{figure}

A particularly instructive feature appears in the neutron star-like configuration at high frequency ($R=6M$, $2M\omega=16$), where the differential cross section exhibits a pronounced  rainbow-like enhancement around $\theta_r\simeq59.6^\circ$. As discussed in Sec.~\ref{sec:results_debye}, this rainbow pattern is already largely contained in the $p=1$ Debye contribution. The CAM decomposition refines this statement by identifying the dominant pole content near $\theta_r$ :  within the $p=1$ Regge-Debye pole sum, the broad-resonance family associated with the zeros of $c^{\rm in}$ provides the main contribution in the neighborhood of the rainbow peak, whereas the surface-wave poles associated with the zeros of $\alpha^{\rm in}$ contribute more smoothly as a function of angle (see Fig.~\ref{fig:CAM_Debye_R6_2Mw16_Rainbow}). By contrast, in the ultracompact regime ($R=2.26M$) the corresponding rainbow angle occurs beyond $\theta=180^\circ$, so that the forward/intermediate-angle spectra considered here do not display an analogous rainbow feature.

\section{Conclusions and outlook}
\label{sec:conclusion}

We have extended the complex angular momentum analysis of scattering by horizonless compact objects by deriving an exact Debye-series expansion of the scattering matrix and by applying CAM techniques separately to each Debye contribution. This Debye-series formulation isolates the direct surface reflection term from the sequence of contributions involving transmission through the surface and interior propagation, and it provides a natural trajectory interpretation: the leading term describes reflection at the surface, whereas higher-order terms correspond to waves that enter the object and return to infinity after successive surface-center-surface traversals, with partial internal reflections at the surface.

The same Debye-series framework leads to a corresponding Regge-Debye pole spectrum in the complex angular-momentum plane. Two complementary sources contribute to this pole content. Zeros of the surface coefficient $\alpha^{\rm in}_{\lambda-1/2}(\omega)$ define a surface-wave family, while zeros of the interior coefficient $c^{\rm in}_{\lambda-1/2}(\omega)$ define an interior-resonance sector. For neutron star-like compactness $R>3M$, this interior sector appears as a broad-resonance branch. In the ultracompact regime $R<3M$, the interior-resonance sector separates further into a broad-resonance branch and a narrow-resonance branch, with the latter lying closer to the real axis and reflecting long-lived, quasitrapped behavior.

We have also reconstructed scattering observables from the Debye partial-wave contributions and assessed the convergence of the Debye reconstruction across frequencies and angles. For the neutron star-like configuration, small-angle scattering is dominated by the direct term, while intermediate and large angles receive substantial contributions from the first interior-transmission term and from interference between several Debye orders. At intermediate frequency only a few Debye orders are required, whereas at higher frequency additional Debye terms are needed to recover the fine oscillatory structure, particularly at large angles. For the ultracompact configuration, higher Debye orders are strongly suppressed in the regimes considered here, and a small number of terms already provides an accurate reconstruction over a broad angular range.

Applying CAM techniques to each Debye amplitude yields exact CAM representations in which every $f^{(p)}$ is written as a Regge-Debye pole sum together with a background integral and a branch-cut contribution, with the principal-value prescription fixed by the physical sheet. These CAM reconstructions agree quantitatively with the corresponding Debye partial-wave results. The CAM decomposition also highlights a qualitative difference between neutron star-like and ultracompact configurations. In the neutron star-like case, the surface-wave pole contribution and the branch-cut contribution are comparable over wide angular ranges; their sum already reproduces the dominant large-angle behavior, including the approach to the backward/glory region. At intermediate and small angles, quantitative agreement requires the additional contribution from the broad-resonance pole sector. In practice, accurate reconstructions at small angles require summing many surface-wave poles. In the computations presented here we retained $N_{\rm S-W}=50$ surface wave poles. By contrast, in the ultracompact regime and over the angular ranges displayed, the Debye amplitudes are dominated by the Regge-Debye pole sums, while the background and branch-cut sectors are strongly suppressed. The split of the interior-resonance sector into broad-resonance and narrow-resonance branches then provides a clear separation between broad and long-lived resonance content.

The Debye-CAM formulation also sharpens the interpretation of the  rainbow-like enhancement in the neutron star-like high-frequency regime. We find that this enhancement is already present in the first interior-transmission contribution and that, within the corresponding pole sum, the broad-resonance family provides the dominant contribution in the neighborhood of the rainbow angle, while the surface-wave family varies more smoothly with $\theta$. This establishes a direct mapping between prominent angular features of the cross section and the internal physics of the compact body, mediated by the specific interior-sensitive Regge-Debye pole family.

It is useful to contrast this behavior with the role played by the Debye series in the classical CAM treatment of Mie scattering~\cite{Nussenzveig:2006}. In that setting, the Debye series is often used to recast the scattering amplitude in a way that effectively bypasses the broad-resonance sector and accelerates convergence. In the compact-object problem studied here, by contrast, the Debye-series expansion does not remove the broad-resonance poles; instead, it separates exactly the surface-associated and interior-associated pole content, thereby allowing a direct attribution of angular features to distinct contributions. A further difference is that, whereas in the classical setting rainbow structures are commonly associated with higher Debye orders ($p=2$), the compact-object problem already exhibits a dominant  rainbow-like signature in the first interior-transmission contribution ($p=1$), reflecting the structure of the curved-spacetime effective potential and the interplay between transmission at the surface and interior propagation.

Several extensions follow naturally from the present work. A natural extension on the theoretical side would be a semiclassical analysis of the Regge-Debye spectrum. In particular, it would be desirable to derive WKB-type asymptotic formulas for the different pole families identified here. In the neutron star-like regime ($R>3M$), such an analysis should account separately for the surface-wave and broad-resonance branches, whereas in the ultracompact regime ($R<3M$) three distinct pole families are expected. Establishing these asymptotic descriptions would provide a deeper semiclassical interpretation of the Regge-Debye spectrum and of its role in shaping the scattering observables. On the phenomenological side, it is natural to generalize the present scalar-field analysis to electromagnetic and gravitational perturbations, and to more realistic interior models (e.g., polytropic equations of state; see Ref.~\cite{Stratton:2019deq}). In that broader setting, the Debye-CAM framework may offer a systematic route to isolating interior-sensitive signatures in strong-field scattering, and hence to identifying robust scattering fingerprints of compact bodies.

\section*{Acknowledgments}

We thank Sam R. Dolan for helpful suggestions and constructive feedback, which improved this work. We also thank the anonymous referees for their valuable comments and suggestions, which helped clarify some points and improve the presentation of the paper.

\section*{Data availability}

The data are not publicly available. The data are available from the author upon reasonable request.

\bibliography{Debye_scattering_Compact_Objects}

\end{document}